\documentclass[twocolumn,amsmath,amssymb,prl,superscriptaddress]{revtex4-1}
\usepackage[]{graphicx}
\usepackage{bm}
\usepackage{xcolor}
\usepackage{booktabs}
\usepackage{dcolumn}

\usepackage{url} % used for data availability section
\def\ket#1{\mbox{\boldmath $#1$}} % used in method section
\newcommand{\bracket}[1]{\left\langle #1 \right\rangle}

\usepackage{natbib} 
\setcitestyle{super}

\AtBeginDocument{
\heavyrulewidth=.08em
\lightrulewidth=.05em
\cmidrulewidth=.03em
\belowrulesep=.65ex
\belowbottomsep=0pt
\aboverulesep=.4ex
\abovetopsep=0pt
\cmidrulesep=\doublerulesep
\cmidrulekern=.5em
\defaultaddspace=.5em
}

 \usepackage{xcolor}  % Coloured text etc.
\begin{document}
%\title{\textbf{Agnostic inference of latent opinion groups in free-response surveys}}
\title{\textbf{Democratic classification of free-format survey responses with a network-based framework}}
%\title{\textbf{Graph-based classification of opinions in large-scale surveys}}

\author{Tatsuro Kawamoto}
\affiliation{Artificial Intelligence Research Center, National Institute of Advanced Industrial Science and Technology, 2-3-26 Aomi, Koto-ku, Tokyo, Japan}
\author{Takaaki Aoki}
\affiliation{Faculty of Education, Kagawa university, Takamatsu, 760-8521, Japan}

\date{\today}
\begin{abstract}
Social surveys have been widely used as a method of obtaining public opinion. Sometimes it is more ideal to collect opinions by presenting questions in free-response formats than in multiple-choice formats. Despite their advantages, free-response questions are rarely used in practice because they usually require manual analysis. Therefore, classification of free-format texts can present a formidable task in large-scale surveys and can be influenced by the interpretations of analysts. In this study, we propose a network-based survey framework in which responses are automatically classified in a statistically principled manner. This can be achieved because, in addition to the texts, similarities among responses are also assessed by each respondent. We demonstrate our approach using a poll on the 2016 US presidential election and a survey taken by graduates of a particular university. The proposed approach helps analysts interpret the underlying semantics of responses in large-scale surveys.
\end{abstract}
 
\maketitle

\section{Introduction}
%\textcolor{blue}{When you are asked to respond to a survey, e.g., about a certain political issue, how would you explain your opinion? Your answer may not be expressed by a single word or phrase. 
%Likely, many other respondents also explain in a way that can hardly be represented by stereotype response. 
%On the other hand, the goal of the survey would be to extract several ideas that the respondents have in common, which can be represented by simple sentences. 
%This is a general issue that survey analyses have. }
A survey is a method of collecting ideas, opinions, and other information from people. They are widely used in many contexts, including social research, patient feedback for medical care, customer survey of products and services, and opinion polls for elections and government policies. 
Surveys questions are typically presented in multiple-choice questions (i.e. closed-ended questions) or free-response questions (i.e. open-ended questions).

A question like ``Are you a scientist?” can be answered with ``yes" or ``no", and therefore can be easily considered in a multiple-choice format. 
However, some questions must be asked in a free-response format \cite{Kahn:1957tc, Schuman:1987cv}. For example,  a question such as ``Why did you become a scientist?” cannot be easily considered in a multiple-choice format. To create a list of multiple choices, an analyst must predict several representative responses. This task can be quite challenging because possible major responses may be missing from the list \cite{Schuman:1996tb,RePass:1971vj,KelleyJr:2014ti}. In such a situation, the outcome of the survey can be significantly influenced by the decision of an analyst. Surveys with free-response questions are free from such difficulties and limitations, directly revealing the logic and thought processes of respondents \cite{Geer:1988kl, Singleton:2017ud}.

Despite their advantages, surveys rarely include free-response questions because they require manual analysis. To perform statistical analyses, the various opinions collected in the form of text data must be classified into a manageable number of categories without introducing any loss in their intrinsic variety and reflecting underlying semantic relations among responses. This is termed {\it coding} \cite{Singleton:2017ud} and is considered a significant scientific issue. This task is subjective and time-consuming even for a trained analyst \cite{Schuman:1966uz, Schuman:1996tb, Lombard:2002wv} and is particularly complicated in the case of large-scale surveys. 
In fact, in the literature \cite{Giddens:2013tz}, the limitations of free-response surveys are viewed as follows: 
\begin{quote}
\textit{`` ... the lack of standardization means that responses are likely to be more difficult to compare statistically, and this limits attempts to draw general conclusions from the study."}
\end{quote}

\begin{figure*}[t!]
 \begin{center}
   \includegraphics[width=\linewidth]{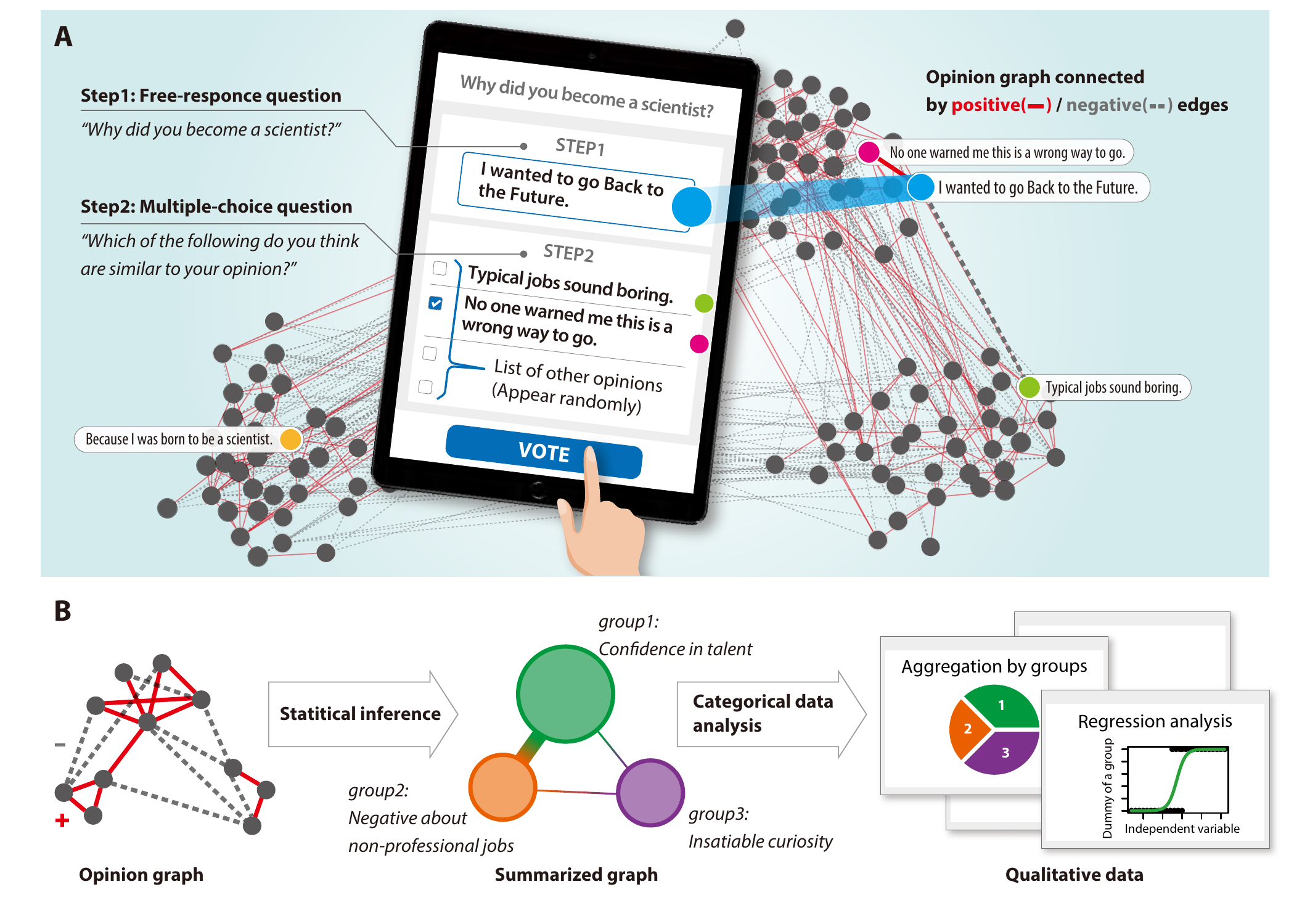}
 \end{center}
 \caption{
 \textbf{Schematic illustration of our network-based approach. }
 {\bf a,} Data collection process in the framework. In Step 1, a respondent inputs his/her own opinion as a response to a free-response question. In Step 2, the respondent refers to a few randomly selected opinions input by other respondents and responds to whether or not he/she feels that these opinions are similar to his/her opinion. This process generates an opinion graph in which a vertex represents a respondent and the corresponding opinion, and an edge represents the relationship between the pair of opinions that it connects. 
 {\bf b,} Procedure of the analysis. Given an opinion graph, the opinions are categorized via graph clustering; this amounts to constructing a small network in which each super vertex represents an opinion group. The opinion groups are then used for categorical data analysis to derive qualitative results. 
 }
\label{FigIllustration}
\end{figure*}

Here we propose a network-based (or graph-based) approach that helps analysts perform coding by automatically classifying responses in a statistically principled manner. This is illustrated in Fig.~\ref{FigIllustration}a. In Step 1, for a given free-response question, the respondent inputs his/her own response. In Step 2, a small subset of the responses input by other respondents is randomly drawn and presented to the respondent. The respondent then selects those responses that he/she feels are similar to his/her own response, if any. This is similar to the manner in which ``likes'' are input in social network services.

After the above procedure has been executed by many respondents, we obtain a dataset that we call the {\it opinion graph} (schematically depicted in the far left-hand side of Fig.~\ref{FigIllustration}b). The vertices denote the responses and the edges encode the ``positive'' (chosen as similar by one of the respondents) or ``negative'' (presented to a respondent, but not chosen) relations among them. In other words, this is a network in which the measurement process is taken into account \cite{Aicher2015,Newman2018,PeixotoPRX2018}. When there is a consensus among the respondents, the resulting network exhibits a modular structure in which vertices within a group are likely to be connected via positive edges (an assortative structure), and vertices in different groups are likely to be connected via negative edges (a disassortative structure). We statistically infer such a modular structure using a graph clustering algorithm, as described in the Materials and Methods. Therefore, in the method that we propose, classification is performed by an algorithm that inputs respondent choices rather than by human analysts. To assign a label to a group, we extract keywords from several typical opinions from that group (middle of Fig.~\ref{FigIllustration}b). We can then perform categorical data analysis for those opinion groups analogously to the case of a survey with multiple-choice questions (far right-hand side of Fig.~\ref{FigIllustration}b).

Estimating the number of groups is also an important problem. For example, the network may not exhibit a clear modular structure. This occurs when the opinions do not exhibit any obvious separation of ideas. An advantage of our approach is that we can assess the statistical significance among opinion groups.

To demonstrate our network-based framework, we implemented an online-survey system. We discuss the results we obtained from two surveys in the following.

\section{Results}

\noindent\textbf{Poll on the 2016 US presidential election}\\ 
The first application of our framework is a poll on the presidential election of the United States in 2016. A total of 117 responses were collected from Oct. 1st to Nov. 8th (election day) 2016. Although the respondents were not restricted to any particular group, a majority of opinions were collected from people who visited the University of Nevada, Las Vegas on 18 and 19 October -- the day before and the day of the final presidential debate. (Note that the University of Nevada, Las Vegas was the venue of this debate.)

\begin{figure*}[!t]
 \begin{center}
   \includegraphics[width=0.95 \linewidth]{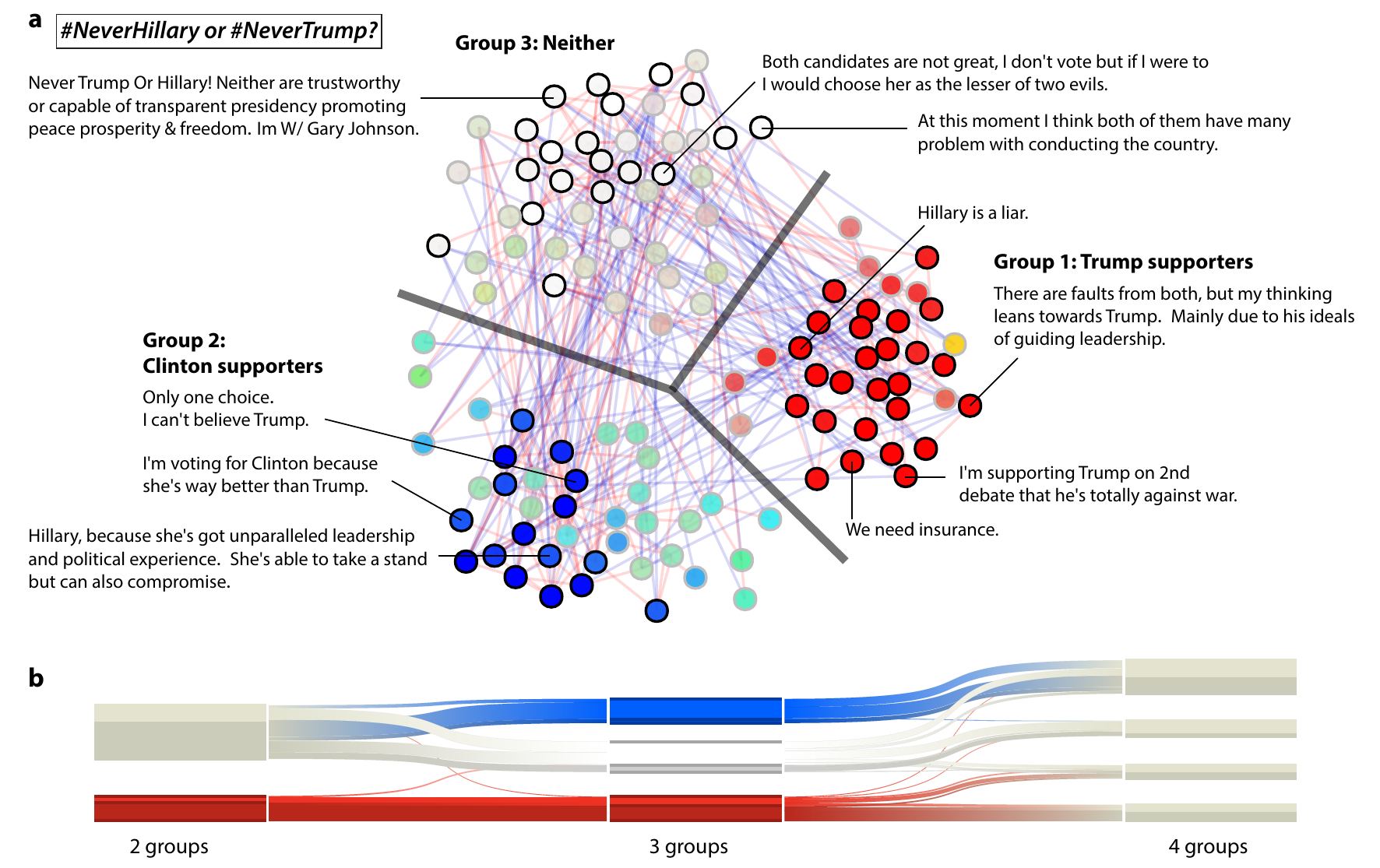}
 \end{center}
 \caption{ {\bf Results of the poll on the 2016 US presidential election.}
 		{\bf a,} Opinion graph for the question ``\#NeverHillary or \#NeverTrump?'' 
		{\bf b,} Alluvial diagram depicting the flow among the partitions with two, three and four groups.
 		}
 \label{FigUSPresident}
\end{figure*}

The question we asked was ``\#NeverHillary or \#NeverTrump?'' 
As shown in Fig.~\ref{FigUSPresident}a, we could extract not only the groups of opinions expressed by the supporters of Donald Trump (group 1) and Hillary Clinton (group 2) but also a group of opinions expressed by those who supported neither (group 3). In Fig.~\ref{FigUSPresident}a, red and blue edges denote the positive and negative edges, respectively. Here, we use a colour gradient ranging between red (group 1), blue (group 2), and white (group 3) to express estimated assignment probabilities ($(p_{1}, p_{2}, p_{3})$) with respect to those three groups. The colour indicates the proximity to the three group; that is, more red means closer to the red group, more blue means closer to the blue group, and more white (a paler shade) means closer to the white group. 
In addition, vertices that are selectively localized within a single group ($\max p_{i} \ge 0.9$) are represented by deep red, deep blue, and pure white and have black borders; these vertices can be regarded as the typical opinions of a group.

Note that whether a respondent supports either of the candidates may not be expressed explicitly in a response. For example, the dataset contains responses such as  `We need insurance' and `Gary Johnson'.
Several such responses can be appropriately classified based on prior knowledge of the candidates. However, the distinction between Clinton supporters and people who support neither are often unclear, whereas Trump supporters can usually be easily distinguished. In such a situation, classification would be extremely difficult if we were to classify opinions using only natural language processing, unless sufficient domain knowledge of the subjects of the survey were incorporated. Instead, what we should really believe in is the decisions made by each individual, that is, the opinion graph.

The alluvial diagram \cite{Rosvall:2010hv} in Fig.~\ref{FigUSPresident}b depicts flows of group assignments as we increase the number of groups. A bundle represents a set of responses that are classified as the same group, and the diagram depicts the flow of group assignments as the number of groups changes. An alluvial diagram is an extension of a Sankey diagram. The diagram expresses the significance of group assignments with colour shading: A bundle of vertices that are selectively localized in a single group with $\max p_i \ge 0.9$ is represented by a deep colour, while others are represented by paler colours. The groups that can be interpreted as Trump supporters, Clinton supporters, and neither correspond to red, blue, and grey, respectively.

Alluvial diagrams provide a useful tool to visually determine the appropriate number of groups (the model selection) \cite{Kawamoto_modBIX} to be used for a given survey. In the present case, it can be observed that groups 2 and 3 form a single group in the case of bipartition, and they split in the case of tripartition. When the network is partitioned into four groups, the responses in deep colours no longer split. Moreover, a significant fraction of vertices in different groups merge, and as a result, the hierarchical structure no longer exists. For these reasons, we selected the partition with three groups as the final result. Further details of the assessment of the proper number of groups are provided in Supplementary Materials.

\bigskip
\noindent\textbf{Social survey of the faculty of education in a particular university} \\
As the second application of our framework, we conducted a survey focussing on the graduates of the Faculty of Education of a particular university.
The questions we asked are the following: `What is your career?' (Q1); `What reason made you choose your current job?' (Q2); `What is the most valuable experience that you will take away with you from your time at the university?' (Q3). 
We announced the survey via postal mail to people who graduated between 2000 and 2016, and we received 258 responses. (The response rate was $\sim$10\%.) 

\begin{figure*}[!t]
 \begin{center}
   \includegraphics[width=\linewidth]{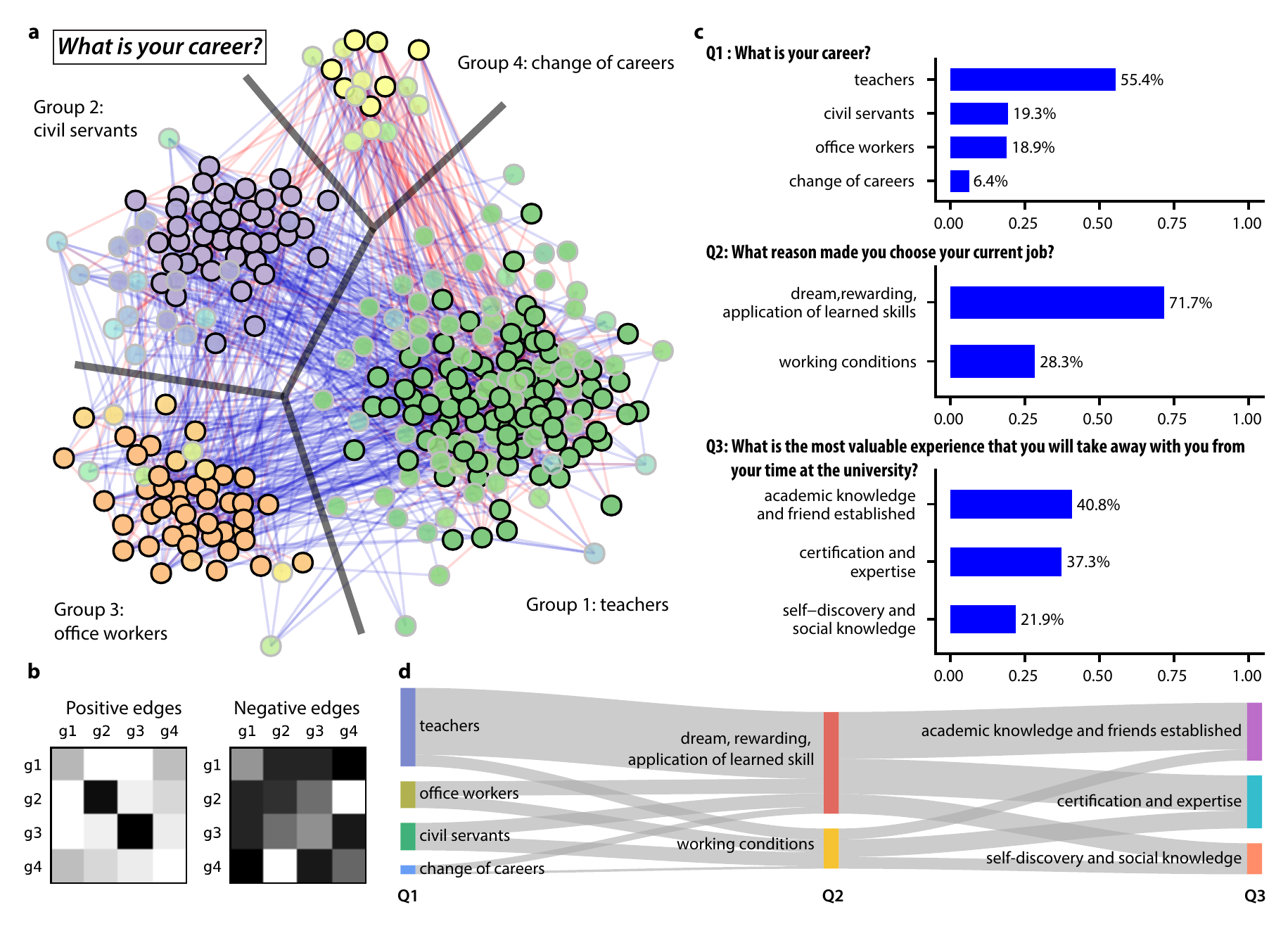}
 \end{center}
 \caption{
{\bf  Results of the survey focussing on graduates of the Faculty of Education.}
{\bf a,} The opinion graph for Q1 -- `What is your career?'
{\bf b,} The connection patterns between/within the groups for Q1. 
{\bf c,} Bar charts of the populations of the opinion groups for Q1, Q2 and Q3. 
{\bf d,} Sankey diagram depicting the flow of respondents among the opinion groups for Q1, Q2 and Q3.
 }
\label{FigKagawa}
\end{figure*}

Figure~\ref{FigKagawa}a presents the opinion graph for Q1 -- `What is your career?' 
The network is plotted in the same manner as that in Fig.~\ref{FigUSPresident}a. 
To determine the appropriate number of groups, we again used the alluvial diagram, as in the case of Fig.~\ref{FigUSPresident}b, including the criteria based on cross-validation estimates of edge-prediction errors (see Supplementary Material for details).

Figure~\ref{FigKagawa}b depicts the structures that the algorithm learned with respect to the positive and negative edges. 
It is observed that except for group 4, the positive edges exhibit an assortative structure, as intended. 
However, it must be noted that negative edges exhibit a structure that is more complicated than a disassortative structure. Therefore, it is evident from our result that the negative edges must not be strictly interpreted as dissimilar relationships.

Once the responses have been classified, we can analyse the categorical data using standard tools. The bar chart for each question is presented in Fig.~\ref{FigKagawa}c. Among the four groups that emerged from the responses to Q1, `teachers' (group 1) has a clear majority; over half of the respondents are assigned to this group. This is reasonable because all respondents are graduates of the Faculty of Education. The other three groups include `civil servants' (group 2), `office workers' in private firms (group 3), and `change of careers' (group 4). (It must be noted here that only group 4 is of a carrier path, instead of a name of the job.) These groups reflect a similarity that the graduates recognize, which is distinct from simple similarities between texts or the standard classification. The respondents would have responded based on their own experiences, the education they have received, career paths of their friends, parents, neighbours, and so on.

Two groups emerged from the responses to Q2 (career motivation): `dream, rewarding, application of learned skills' and `working conditions'. The responses to Q3 (valuable experience) were classified into three groups: `academic knowledge and friendships established', `certification and expertise', and `self-discovery and social knowledge'. Although some of the typical responses within a given group may appear quite distinct at the first glance, these responses form a peer group with the same statistical tendencies of positive and negative edges.

We can again confirm that the opinion graphs reflect a similarity based on the underlying semantics, instead of simple text similarities. 
For example, in the responses of Q1 (``What is your career?''), a response ``a school janitor of a primary school'' is positively connected to ``a municipal officer,'' while it is negatively connected to ``a primary school teacher.''
%If we simply assess the similarity between texts, it appears that the signs of the edges should be opposite. However, it is not a mistake. In fact, considering the careers of these graduates, it actually makes sense. 
As another example, in Q3 (``What is the most valuable experience that you will take away with you from your time at the university?''), it can be observed that a response ``I found what I really wish to do in my life.'' is positively connected to ``I met many people and I could have time to face myself.'' 
%Although we never know the exact logic behind, it is evident that this response is semantic and is more than just a similarity between texts. 

To confirm the meaningfulness of the identified groups, we compared them with human-based classifications in which human analysts read all the responses and classified them into several categories using their own interpretations. We observed that the identified groups are in good agreement with classifications performed by some of the analysts, in the sense of normalised mutual information (NMI) \cite{NMI} and adjusted Rand index (ARI) \cite{Randindex,ARI}. The results of the comparison are presented in Supplementary Materials.

%The classification in Q1 is clearly affected by the recognition about the careers shared among the respondents, and is distinct from the standard classifications of occupations used in statistics. 
%For example, to be a school teacher or a civil servant, one has to pass the corresponding deployment examination. 
%On the other hand, the job hunting in private firms starts in advance of these examinations. 
%Finally, \textit{Job changes} is probably regarded as none of the above cases, and is also distinct among the respondents. 
%If this survey were done for another faculty or another social group, different classification may arise, reflecting on the background of the respondents. 

The Sankey diagram in Fig.~\ref{FigKagawa}d depicts the flows of respondents among the opinion groups for Q1, Q2, and Q3. It appears that the respondents in the group `teachers' for Q1 mostly flow into the group `dream, rewarding, application of learned skills' for Q2, and a relatively large fraction of them flow into the group `academic knowledge and friendships established' for Q3. This observation suggests some correlations between the factors queried in Q1, Q2, and Q3. This can be further confirmed by performing regression analysis. 
%\textcolor{green}{The regression analysis that we employed is described in Supplementary Materials. [Omit this, maybe?]}
A more detailed analysis of the above surveys can be found in Supplementary Materials.

\section{Discussion}
To overcome the problems that limit the usefulness of free-response questions in large-scale surveys, we proposed a network-based survey framework. The advantages of the present approach are discussed in the following.

\begin{description}
\item[ \textit{Tractability and scalability} ] The conversion of a large set of raw text data into an opinion graph makes the analysis much more tractable and facilitates coding for numerous responses.
\end{description}

One classical prescription to facilitate coding is to conduct surveys twice, first conducting a small-scale survey using free-response questions to determine the major responses and then conducting a large-scale survey using multiple-choice questions whose possible responses are based on the major opinions obtained previously. However, not only does this two-step survey require additional effort, but there still exists a possibility of missing major responses in the large population. Another approach that can be used for coding is to perform some natural language processing (NLP) \cite{Simon:2004kc, Hopkins:2010tu, Roberts:2014es}. However, such an approach is effective only when one has a dataset that contains the appropriate domain knowledge of the survey. 
To confirm if this is the case for the datasets, we conducted the NLP-based classifications in the Supplementary Materials. 
In contrast, once the opinion graph is constructed in our framework, the responses are automatically classified without additional costs or the need for any prior knowledge. This makes it possible to conduct a large-scale survey with free-response questions in an efficient manner.

\begin{description}
\item[ \textit{Statistically principled and democratic} ] The classification reflects the collective evaluations made by the respondents themselves, instead of the decisions by a small number of analysts. Thus, the result is not subjective or completely objective, but is democratic. 
\end{description}
The difficulty in drawing general conclusions is often explained as a weakness of free-response surveys \cite{Giddens:2013tz}. By contrast, with our framework, it is possible to draw general conclusions from free-response questions. This is possible because 
(1) the diversity of responses is assessed in a statistically principled manner in terms of the network configuration, and  
(2) the network is provided by the subjective evaluations of relationships among responses accumulated from the respondents. 
%In this sense, the present approach can be regarded as unsupervised crowdsourcing.
In contrast, as often done in the standard crowdsourcing \cite{Benoit_2016,Lind_2017,Jacobson_2018}, we could let employed workers perform Step 2 as hypothetical respondents. 
In that case, however, the workers' coding abilities would play a critical role. 
We would then no longer have the same objective that we consider in this paper.

%% Factor analysis
Factor analysis is a method frequently used in survey analysis, particularly in the field of psychology. 
Using the responses to a series of multiple-choice questions, the analysts are able to extract a few factors that characterize those responses from various aspects. 
It should be noted, however, that the factor analysis suffers from the same problems that the multiple-choice questions have in general. 
%, because the factor analysis is performed on top of the responses to the multiple-choice questions after all. 

%% Detectability limit
While this framework paves the way for a number of investigations using free-response surveys, it also includes novel problems concerning both survey design and statistical theory of networks. For example, there is a trade-off between the effort made by the respondents and the degree to which the groups of responses are resolved. In general, the sparser the network, the more information-theoretically difficult it will be to identify the appropriate opinion groups. If the framework requires a respondent to consider a large number of responses, the resulting opinion graph will be dense. Evidently, demanding too much effort from respondents is unrealistic. On the other hand, if each respondent only considers a small number of responses, the resulting network will be sparse. This trade-off is related to the certainty of consensus behind the responses; a set of clearly distinguishable responses can be detected even in a very sparse network, while a dense network is needed to detect subtle differences between opinion groups. The limit of distinguishability of a modular structure of sparse networks is called the detectability limit or the detectability threshold \cite{Decelle2011PRE,MooreReview2017} in statistical theory of networks. The difficulty associated with coding in large-scale surveys is directly related to this limit.

%% Influence by other respondents
A possible bias in Step 2 should also be mentioned. 
The respondents may be influenced by the responses of others. 
When this point is a critical issue, we can maintain independence among responses by simply hiding Step 2 until Step 1 is completed according to the procedure depicted in Fig.~\ref{FigIllustration}a. 
However, a more important issue in free-response surveys is the problem of 'frame of reference' \cite{Schuman:1966uz,Schuman:1987cv}, i.e. the risk of failing to collect sufficiently diverse opinions while keeping the coherence among them.  %that we cannot collect enough diversity of opinions while keeping the coherence among them. 
Our framework is more suitable for the type of survey in which the respondents are expected to be well-informed \cite{Fishkin2011} and speculate deeply in reference to the opinions of others. %as referring to various opinions. 
In such a situation, displaying other responses can be considered beneficial rather than detrimental. 
%For these reasons, we permitted respondents to refer to the responses of other respondents. 

Although evaluating the diversity of responses is problematic in many senses, it is always needed whenever a survey seeks information from the public. We hope that our framework will play a useful role in society and help researchers acquire new knowledge in various fields of science.

\section{Materials and Methods}
\noindent\textbf{Graph clustering based on statistical inference}\\
We perform the partitioning of an opinion graph using graph clustering, which is a component of probabilistic machine learning. This task is also known as community detection, in more restricted cases. Although numerous frameworks and algorithms have been proposed in the literature, in this study, we employ the statistical inference approach. 

A random graph model with a modular structure called the stochastic block model \cite{holland1983stochastic,WangWong87} is a canonical generative model for the inference of a modular structure. 
%Because the edges are labeled, we consider a variant called the labeled stochastic block model. 
The group label is denoted by $\sigma \in \{1,\dots,q\}$, where $q$ is the number of groups, and the set of group assignments is denoted by $\ket{\sigma}$. 
The model parameters, $\ket{\gamma}$, $\ket{\omega}^{+}$ and $\ket{\omega}^{-}$ specify the macroscopic structure of the network: 
$\ket{\gamma}$ is a $q$-dimensional vector that represents the fractions of group sizes, and $\ket{\omega}^{+}$ and $\ket{\omega}^{-}$ represent the $q \times q$ density matrices (also known as the affinity matrices) that determine the connection probabilities within/between groups with respect to the positive and negative edges, respectively. 
An instance of the stochastic block model, i.e. the adjacency matrix $\ket{A}$, is generated as follows. 
For each vertex, we randomly determine the group assignment such that the fraction of the group $\sigma$ is equal to $\gamma_{\sigma}$ on average. 
Then, for all pairs of vertices, we generate edges independently and randomly on the basis of group assignments. 
For example, when vertices $i$ and $j$ have the assignments $\sigma_{i} = \sigma$ and $\sigma_{j} = \sigma^{\prime}$, they are connected by a positive edge with probability $\omega^{+}_{\sigma \sigma^{\prime}}$, connected by a negative edge with probability $\omega^{-}_{\sigma \sigma^{\prime}}$, and not connected with probability $1 - \omega^{+}_{\sigma \sigma^{\prime}} - \omega^{-}_{\sigma \sigma^{\prime}}$. 
(This type of stochastic block model is sometimes referred to as a labelled stochastic block model.) 

It is important to note that we can enforce the sampling of vertices referred by a respondent to be random.
Therefore, the probabilistic nature of the edge generation process assumed in the stochastic block model is guaranteed. 
It is not necessary that the generation process of a real data resembles to that of the model for the stochastic block model to be applicable. 
However, the results will be more reliable when they are similar. 
%In this sense, the stochastic block model is well suited for the partitioning of the opinion graph. 

The likelihood function of the stochastic block model with $N$ vertices is given by 
\begin{align}
&p(\ket{A}, \ket{\sigma} | \ket{\gamma}, \ket{\omega}^{+}, \ket{\omega}^{-}) \notag\\
%&= p(\ket{A} | \ket{\sigma}, \ket{\omega}^{+}, \ket{\omega}^{-}) p(\ket{\sigma} |\ket{\gamma}) \notag\\
&\hspace{10pt}= \prod_{i=1}^{N} \gamma_{\sigma_{i}} \prod_{i<j} 
\biggl[ \left(1 - (\omega^{+}_{\sigma_{i}\sigma_{j}} + \omega^{-}_{\sigma_{i}\sigma_{j}}) \right)^{\delta_{A_{ij},0}} \notag\\
&\hspace{20pt}\times \left( \omega^{+}_{\sigma_{i}\sigma_{j}} \right)^{\delta_{A_{ij},+1}}
\left( \omega^{-}_{\sigma_{i}\sigma_{j}} \right)^{\delta_{A_{ij},-1}} \biggr]. \label{labeledSBMlikelihood}
\end{align}

This stochastic block model assumes that all of the vertices within a given group are statistically equivalent. However, in the opinion graph, even within a group of similar responses, some responses may be more popular than others. This heterogeneity is manifested as a hub structure in the opinion graph, which cannot be modelled by equation (\ref{labeledSBMlikelihood}). 
A variant of the stochastic block model that properly considers a hub structure is called the degree-corrected stochastic block model \cite{KarrerNewman2011}. 
The corresponding likelihood function in this model is given by 
\begin{align}
&p(\ket{A}, \ket{\sigma} | \ket{\gamma}, \ket{\omega}^{+}, \ket{\omega}^{-}) \notag\\
%&= p(\ket{A} | \ket{\sigma}, \ket{\omega}^{+}, \ket{\omega}^{-}) p(\ket{\sigma} |\ket{\gamma}) \notag\\
&\hspace{10pt}= \prod_{i=1}^{N} \gamma_{\sigma_{i}} \prod_{i<j} 
\biggl[ \left(1 - (d^{+}_{i}\omega^{+}_{\sigma_{i}\sigma_{j}}d^{+}_{j} + d^{-}_{i}\omega^{-}_{\sigma_{i}\sigma_{j}}d^{-}_{j}) \right)^{\delta_{A_{ij},0}} \notag\\
&\hspace{20pt}\times \left( d^{+}_{i}\omega^{+}_{\sigma_{i}\sigma_{j}}d^{+}_{j} \right)^{\delta_{A_{ij},+1}}
\left( d^{-}_{i}\omega^{-}_{\sigma_{i}\sigma_{j}}d^{-}_{j} \right)^{\delta_{A_{ij},-1}} \biggr], \label{labeledDCSBMlikelihood}
\end{align}
where $d^{+}_{i}$ and $d^{-}_{i}$ represent the degrees, i.e. the number of neighbours connected via positive and negative edges from vertex $i$, respectively. The degrees of vertices can be readily obtained from $\ket{A}$. (Note also that the density matrices are renormalised compared to those in equation (\ref{labeledSBMlikelihood}), owing to the degree distribution.)

Based on equation (\ref{labeledDCSBMlikelihood}), we infer the group assignment $\sigma_{i}$ of vertex $i$ by solving for its marginal posterior distribution, $p(\sigma_{i} | A, \ket{\gamma}, \ket{\omega}^{+}, \ket{\omega}^{-})$, and the values of the model parameters are determined so that the marginal likelihood $\sum_{\ket{\sigma}} p(A, \ket{\sigma} | \ket{\gamma}, \ket{\omega}^{+}, \ket{\omega}^{-})$ is maximised. 
Unfortunately, precise computations of the marginal posterior distributions and model parameters are computationally demanding. 
A commonly used approximation method  is the expectation-maximisation (EM) algorithm \cite{BishopPRML}, and belief propagation \cite{MezardMontanari2009,Decelle2011PRE,Decelle2011PRL} is known to be an efficient and accurate algorithm for its E-step. 
We implemented this approach for the stochastic block model. 
We note here that other methods such as the Monte Carlo method can also be used, as long as they  are computationally feasible and applicable to the opinion graphs. 

There are three reasons for employing the inference algorithm described above. First, using this algorithm, instead of a single partition that optimizes a certain objective function, we can obtain a degree of certainty as the probability of group assignment for each opinion vertex. 
Second, the properties of the stochastic block model are theoretically well known \cite{Decelle2011PRE,Kawamoto_ADT-Full,AbbleReview2017,MooreReview2017,Peixoto2017tutorial}, and efficient algorithms exist for the sparse case. 
Third, the algorithm can learn an arbitrary connection pattern for each edge type; that is, we do not need to assume that negative edges strictly exhibit a disassortative structure.

Although we considered only positive and negative edges, it is possible to allow more than two labels and make finer distinctions among responses. In principle, the opinion graph would then acquire more information. However, it must be noted that this does not necessarily imply better identification of groups in practice because it might be very difficult to devise an algorithm that can stably and accurately treat opinion graphs with a large number of edge types \cite{Kawamoto_ADT-Lett,Kawamoto_ADT-Full}. 

To determine the appropriate number of groups, we employed leave-one-out cross-validation estimates of prediction errors, which can be evaluated efficiently \cite{Kawamoto_sbmBIX} using belief propagation. Note that we should not expect such estimates to be very accurate. Their role is merely to provide a rough estimate using a mathematically principled approach. For the final determination of the number of groups, we examined the manner in which the network is actually partitioned using the alluvial diagram.

\subsection*{Acknowledgments}
The authors thank Harumi Tokioka and Shigeru Shinomoto for fruitful discussions. The authors are also grateful to Juyong Park and Martin Rosvall for their comments. 
Finally, the authors appreciate all the people who contributed to the poll on the 2016 US presidential election and acknowledge support from the Faculty of Education in Kagawa University and the reunion of the faculty. 
T.K. was supported by JSPS (Japan) KAKENHI grant no. 26011023. T.A. was supported by the Research Institute for Mathematical Sciences, a joint research centre at Kyoto University, and open collaborative research at the National Institute of Informatics (NII) Japan (FY2017). T.K. and T.A. acknowledge financial support from JSPS KAKENHI grant no. 18K18604.

\subsection*{Data  availability}
The network datasets that support the findings of this study are available in a GitHub repository at \url{https://github.com/tatsuro-kawamoto/opinion_graphs}.

%Poll on the 2016 US presidential election: Raw text data of responses are presented in Supplementary Materials. A graph dataset in the GML format, including the raw texts and inferred group labels, is available at \url{https://github.com/tatsuro-kawamoto/opiniongraph_2016USpresident}. Social survey conducted on a faculty of education of a particular Japanese university: Text data that are translated from Japanese to English are partially presented in Supplementary Materials. Additional data and information can be requested from the authors, addressed to the corresponding author. Alluvial diagrams were generated using Alluvial Generator at \url{http://www.mapequation.org/apps/AlluvialGenerator.html}.

\subsection*{Code availability}
The graph clustering code that supports the findings of this study is available in a GitHub repository at \url{https://github.com/tatsuro-kawamoto/graphBIX}.

%\subsection*{Author contributions}
%Both T.K. and T.A. designed the survey framework, analysed the data, and wrote the manuscript. T.K. implemented the online-survey system. T.K. conducted a survey of the poll on the 2016 US presidential election and T.A. mainly conducted a survey focussing on graduates of the Faculty of Education of a particular university.
%
%\subsection*{Conpeting interests}
%The authors declare that they have no competing interests.

\clearpage

%%%%%%%%%% Merge with supplemental materials %%%%%%%%%%
\pagebreak
\widetext
\begin{center}
\textbf{\large Supplemental Material: \\ Democratic summary of public opinions in free-response surveys}
\end{center}
%%%%%%%%%% Merge with supplemental materials %%%%%%%%%%
% Suppl.
\setcounter{figure}{0}
\setcounter{table}{0}
\renewcommand{\thesection}{S\arabic{section}}
\renewcommand{\thefigure}{S\arabic{figure}}
\renewcommand{\thetable}{S\arabic{table}}

\section{Details on the survey framework, opinion graph, and graph clustering}

\subsection{Attachment of labels on the opinion groups}
Attaching labels to identified groups is performed by human analysts by referring to typical responses. 
Typical responses are the responses with highly localised assignment probabilities in each group that are expected to represent the core ideas of the group. 
When analysts can derive one representative idea shared within a group, a single label is assigned to the group. 
Otherwise, multiple labels are assigned to the group. 

\subsection{Step 1 can be skipped}
Free-response questions can be more laborious for respondents to answer  compared to multiple-choice questions. 
However, it is important to note that a respondent is allowed to skip Step 1 in Fig.~1A. 
In other words, for respondents who choose not to input their own responses, our framework naturally functions as a multiple-choice survey. 
Even in this case, when a respondent fails to input his/her own response, a new vertex corresponding to this respondent is added to the opinion graph. The opinion text corresponding to this vertex is not left empty, but instead, an opinion from one of the responses that the respondent chose as similar is used as the opinion text.

\subsection{Initial state of the survey}
At the start of a survey, there are no previously submitted responses, and therefore, some seed responses must be added. 
In the case of the poll on the US presidential election, we selected six responses that we determined to be typical of Clinton supporters and six responses that we determined to be typical of Trump supporters. 
In the case of the survey focussing on the graduates of Faculty of Education, several seed responses were inputted for each question by a faculty sociologist.
Following the identification of groups, we excluded these seed responses from the analysis. 

These seed responses have both positive and negative effects. Analogous to the argument in \cite{Schuman:1987cv}, when the frame of reference \cite{Schuman:1966uz}, i.e., the domain of response, that the respondents have in mind is too narrow, the seed responses widen their frame. On the other hand, these seed responses can induce bias to the result.

\subsection{Randomness of the selection of responses}
Because the opinion texts are copied when respondents fail to input their own responses, identical responses may appear in the dataset, indicating that some responses are more common than others. 
To prevent redundant sampling of responses, instead of using uniform sampling from the dataset in Step 2 of the procedure depicted in Fig.~1A, responses were  uniformly and randomly drawn from the set of \textit{unique responses}. 

%\subsection{Independence of the responses}
%We can maintain independence among responses (i.e., preventing respondents from being influenced by the responses of other respondents) by simply hiding Step 2 until Step 1 is completed according to the procedure depicted in Fig.~1A.
%Therefore, for example, the priming effect is an issue, this approach should be used. 
%However, in a survey of free-response questions, it is often important that the respondents are well-informed \cite{Fishkin2011}, and coherent responses can be obtained by using other responses as a frame of reference \cite{Schuman:1966uz}. 
%In such a situation, displaying other responses can be considered beneficial rather than detrimental. 
%For these reasons, we permitted respondents to refer to the responses of other respondents. 
 
\subsection{Random edge discarding}
It is known that when the number of negative edges is much larger than the number of positive edges, detectability of the modular structure can be adversely affected \cite{Kawamoto_ADT-Lett,Kawamoto_ADT-Full}. 
To avoid this problem in such a case, we randomly discard a fraction of the negative edges so that the number of positive and negative edges are almost balanced. 
More precisely, for ease of implementation, instead of actually removing some edges from the dataset, we ``neutralized'' them by introducing a new edge label that we refer to as ``neutral,'' and edges are discarded by assigning this label. Because the corresponding density matrix, which is denoted as $\ket{\omega}^{0}$, is set to be uniform, neutral edges are equivalent to unobserved edges.

\subsection{Growth of the opinion graph}
In the stochastic block model, it is assumed that all vertices exist at the beginning, and then edges are independently and randomly generated. In contrast, in the present case, vertices are added as edges are generated, i.e., the opinion graph grows. Because the number of vertices that a respondent can refer to is limited, the vertices added earlier are likely to be referred to more often. It is possible to ensure that all opinions are referred to with equal probability. In that case, however, the respondents at a later time would need to refer to more opinions than the respondents at an earlier time. 

The above feature is only for the case when we sample opinions uniformly and randomly. We emphasize that the survey design allows us to control the statistics of the growth of the opinion graph. Therefore, it is interesting to consider a sophisticated survey design and the corresponding statistical model.

\subsection{Determination of the number of groups}
When comparing partitions with different numbers of groups, our selection of the optimal number of groups must consider both the statistical significance and the interpretability of groups. 

\subsubsection{Determination based on the statistical significance}
In terms of statistical inference, selecting the optimal number of groups in a network is a model selection problem. 
The optimal number of groups is generally determined by evaluating a certain criterion. 
To this end, a number of criteria have been proposed in the literature \cite{Fortunato2010,Luxburg2007,Krzakala2013,Peixoto2017tutorial,NewmanReinert2016}. 
In this study, we employ criteria that are based on prediction errors. 

In machine learning, model selection based on a prediction error is commonly used \cite{Akaike1998,Arlot2010}.
Let us first explain model selection in terms of the prediction error in a general context. 
To measure the prediction error in a given dataset, we split the dataset into two parts: a training set and a test set. 
The values of the model parameters are determined, and the hidden variables are inferred from the training set. Predictability is then evaluated using the unseen test set. 
In such a procedure, when the model complexity is too low, the model cannot correctly predict the information of the test set; i.e., the prediction error becomes large. 
On the other hand, if the model complexity is too large and too much information is extracted from the training set, the prediction error again becomes large. 
Therefore,  underfitting and overfitting are both penalized, and the model that most efficiently describes the dataset is selected as the one that minimizes the prediction error. 

If we do not have sufficient data points with respect to each quantity to be predicted, the obtained prediction error is not reliable because it may be sensitive to the manner in which the dataset is split into training and test sets. 
In such a situation, cross-validation \cite{Arlot2010} provides an adequate estimate. 
In cross-validation, we can consider many ways of splitting the dataset into a training set and a test set, and the prediction errors are computed for each of these. 
In the end, all of the prediction errors are averaged. 
When the dataset is split into $K$ sets (one of which is used as a test set), it is referred as ``$K$-fold cross-validation.'' 
When we leave only one element as a test set and use the remaining as a training set, it is referred as ``leave-one-out cross-validation'' (LOOCV). 
In the current context, the dataset is a network, which is a set of edges, and the LOOCV constitutes an evaluation of the predictability of the existence and non-existence of an edge between each hidden vertex pair. 

An advantage fo the cross-validation error is its wide applicability as it does not rely on asymptotic analysis. 
On the other hand, a disadvantage of the cross-validation error is that it can often be computationally demanding. 
However, the LOOCV is an exceptional case. It can be evaluated efficiently \cite{Kawamoto_sbmBIX} using belief propagation \cite{MezardMontanari2009,Decelle2011PRL,Decelle2011PRE}, and the behaviour of the corresponding estimates are theoretically well known for the number of groups in the stochastic block model \cite{Kawamoto_sbmBIX}.
In this paper, we refer to the LOOCV estimates of prediction errors as ``cross-validation errors''.

\subsubsection{Determination based on the interpretability of groups}
In many cases, instead of a single plausible number of groups, the cross-validation errors provide its candidates, considering the error bars of the prediction errors.
In such situations, we must proceed to a finer inspection by evaluating the manner in which the network is actually partitioned.
As presented in the Results section, visualization provided by the alluvial diagram facilitates better insight into the results of statistical inference \cite{Kawamoto_modBIX}. 
We verified whether the candidate partitions constitute a hierarchical structure, the robustness of small groups, and most importantly, whether these partitions are interpretable.

\section{Raw data and detailed analyses of the conducted surveys}
\subsection{Poll on the 2016 US presidential election}
Among 117 responses, 70 were collected from the University of Nevada, LasVegas, and 35 were collected directly through a web application.
The remaining 12 responses were collected from Twitter and an online article and were used as seed responses. 
Independent responses are listed in Table \ref{tbl:USrawdata}, and typical (black) and non-typical (grey) responses are distinguished. 

\begin{figure}[!t]
 \begin{center}
   \includegraphics[width=\textwidth]{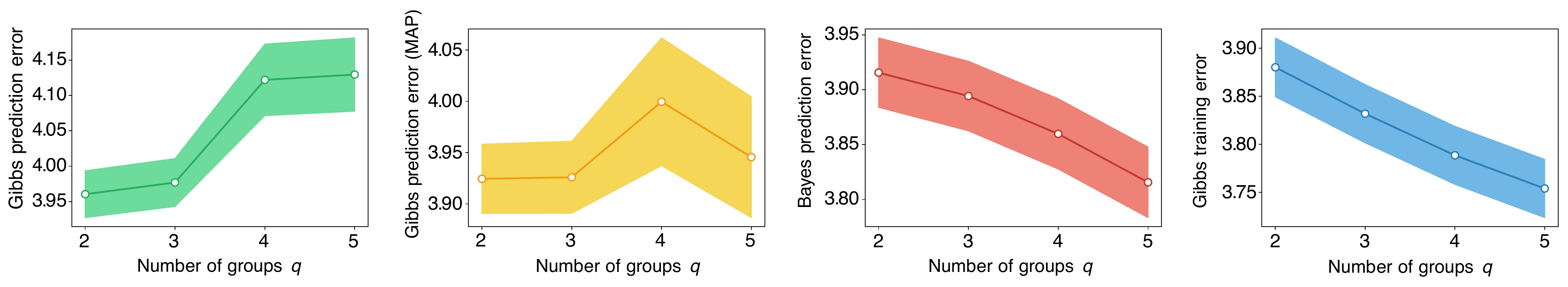}
 \end{center}
 \caption{
 Cross-validation errors for the opinion graph of the poll on the 2016 US presidential election. 
 From left to right: 
 the Gibbs prediction error (green), 
 the MAP estimate of the Gibbs prediction error (yellow), 
 the Bayes prediction error (red), and 
 the Gibbs training error (blue).
 The solid curves and shaded regions represent the mean values and standard errors. 
 Although the prediction errors in the case $q=1$ are not shown, we confirmed that they are much larger than those in the case $q \ge 2$. 
}
\label{fig:LOOCV-US}
\end{figure}

For this data, we demonstrate the validation curves of the cross-validation errors even though we referred directly to the alluvial diagram to determine the number of groups in the main text.
Three types of prediction errors and one type of training error are presented in Fig.~\ref{fig:LOOCV-US}: the Gibbs prediction error (green), the MAP estimate of the Gibbs prediction error (yellow), the Bayes prediction error (red), and the Gibbs training error (blue). 
The definitions of these errors and detailed descriptions of their properties are given in \cite{Kawamoto_sbmBIX}. 
The Gibbs training error is computed only as a reference and cannot be used to determine the number of groups. 
The solid curves in Fig.~\ref{fig:LOOCV-US} indicate the mean values, and the shaded regions indicate standard errors.

First, we confirmed that the values of the prediction errors in the case $q=1$ are much larger than those with $q \ge 2$.
Therefore, the opinion graph is likely to include at least a significant partition. 
From the obtained validation curves, $q=2$ and $q=3$ produce smaller Gibbs prediction errors than the other cases.
As we observed in the alluvial diagram presented in the main text, the partition into two groups roughly distinguishes between Trump supporters and non-Trump supporters, and the latter further splits into the groups of Clinton supporters (group 2) and neither (group 3) in the case $q=3$. 
Such an interpretable hierarchical structure is observed only up to $q=3$. In this paper, we focus on the partition into three groups.

\begin{table}[th!]
\caption{Raw data (responses) for the poll on the 2016 US presidential election and the inferred labels.}
 \begin{tabular}{p{5em} p{3em} p{2em} p{38em}}
\toprule
Label & typical & count & responses \\
\midrule
Trump &Yes &6 & \textit{Hillary Clinton doesn't care about BlackLives She only cares about BlackVotes.} \footnotemark[1]\\
supporter &Yes &3 & \textit{At this moment I think both of them have many problem with conducting the country.}\\
(group 1) &Yes &3 & \textit{She’s a criminal. She does not deserve to hold office. Trump just has a loud mouth. Hillary will destroy the country.} \footnotemark[2]\\
 &Yes &2 & \textit{There are faults from both, but my thinking leans towards Trump. Mainly due to his ideals of guiding leadership.}\\
 &Yes &2 & \textit{I believe that both are incapable of running this country.}\\
 &Yes &2 & \textit{She's a liar and should buy Glennbeck's new book.} \footnotemark[3]\\
 &Yes &1 & \textit{Both options are terrible, break the two party system!}\\
 &Yes &1 & \textit{Calling millions of Americans a basket of deplorables is frightening, hearing this makes her Negroponte endorsement make sense.} \footnotemark[4]\\
 &Yes &1 & \textit{Hillary is a liar. Period.}\\
 &Yes &1 & \textit{Hillary should go to jail.}\\
 &Yes &1 & \textit{I'm supporting Trump on 2nd debate that he's totally against war.}\\
 &Yes &1 & \textit{NeverTrump bc public persona of cruelty and predatory business practices, and a campaign conjuring xenophobia to answer with empty slogans.}\\
 &Yes &1 & \textit{She's a traitorous liar.}\\
 &Yes &1 & \textit{We need insurance.}\\
 &\textcolor{gray}{No} &\textcolor{gray}{2} & \textcolor{gray}{\textit{At this moment I think both of them have many problem with conducting the country.}}\\
 &\textcolor{gray}{No} &\textcolor{gray}{2} & \textcolor{gray}{\textit{Hillary is a liar. Period.}}\\
 &\textcolor{gray}{No} &\textcolor{gray}{1} & \textcolor{gray}{\textit{``I support Donald Trump because a wall would have kept Mark Sanchez from taking my job.''}} \footnotemark[5]\\
 &\textcolor{gray}{No} &\textcolor{gray}{1} & \textcolor{gray}{\textit{Choosing between a liar and a lunatic is hard.}}\\
 &\textcolor{gray}{No} &\textcolor{gray}{1} & \textcolor{gray}{\textit{Hillary Clinton doesn't care about BlackLives She only cares about BlackVotes.}} \footnotemark[1]\\
 &\textcolor{gray}{No} &\textcolor{gray}{1} & \textcolor{gray}{\textit{Trump has great ideas. He's a very brilliant man.}}\\
 &\textcolor{gray}{No} &\textcolor{gray}{1} & \textcolor{gray}{\textit{Trump makes fun of special needs individuals which is very unethical and rude.}}\\
 \midrule
Clinton &Yes &6 & \textit{Hillary has such an impressive history of working for us all!} \footnotemark[6]\\
supporter (group 2) &Yes &3 & \textit{ImWithHer because she cared enough to prepare. She's smart. Ready. And passionate about details because they matter.} \footnotemark[7]\\
 &Yes &1 & \textit{Hillary knows how it works.}\\
 &Yes &1 & \textit{Hillary, because she's got unparalleled leadership and political experience. She's able to take a stand but can also compromise.}\\
 &Yes &1 & \textit{I feel America needs a leader that believes in America and its dream.}\\
 &Yes &1 & \textit{I'm voting for Clinton because she's way better than Trump.}\\
 &Yes &1 & \textit{Only one choice. I can't believe Trump.}\\
 &Yes &1 & \textit{She believes America is already great and will improve it even more!} \footnotemark[8]\\
 &Yes &1 & \textit{Trump would be a disaster to the Japan-Us alliance.}\\
 &\textcolor{gray}{No} &\textcolor{gray}{1} & \textcolor{gray}{\textit{Because she's a better choice than Donald Trump and isn't hateful and racist.}}\\
 &\textcolor{gray}{No} &\textcolor{gray}{1} & \textcolor{gray}{\textit{Both not the best candidates, but \#Imwithher.}}\\
 &\textcolor{gray}{No} &\textcolor{gray}{2} & \textcolor{gray}{\textit{Donald Trump is a sham. He is only interested in his own popularity and power, and is a sexist, racist liar.}}\\
 &\textcolor{gray}{No} &\textcolor{gray}{1} & \textcolor{gray}{\textit{Gary Johnson.}}\\
 &\textcolor{gray}{No} &\textcolor{gray}{5} & \textcolor{gray}{\textit{Hillary has such an impressive history of working for us all!}} \footnotemark[6]\\
 &\textcolor{gray}{No} &\textcolor{gray}{1} & \textcolor{gray}{\textit{I am not a HillaryClinton fan, but I hope she's okay.}} \footnotemark[9]\\
 &\textcolor{gray}{No} &\textcolor{gray}{6} & \textcolor{gray}{\textit{Clinton,No,ImWithHer because she cared enough to prepare. She's smart. Ready. And passionate about details because they matter.}} \footnotemark[7]\\
 &\textcolor{gray}{No} &\textcolor{gray}{1} & \textcolor{gray}{\textit{She believes America is already great and will improve it even more!}} \footnotemark[8]\\
 &\textcolor{gray}{No} &\textcolor{gray}{2} & \textcolor{gray}{\textit{She has flaws but she is still the more informed candidate.}}\\
 &\textcolor{gray}{No} &\textcolor{gray}{1} & \textcolor{gray}{\textit{She is current and he is horrible. I disagree with almost all her policies. Never either.}}\\
 &\textcolor{gray}{No} &\textcolor{gray}{1} & \textcolor{gray}{\textit{Trump would be a disaster to the Japan-Us alliance.}}\\
 &\textcolor{gray}{No} &\textcolor{gray}{1} & \textcolor{gray}{\textit{While Hillary Clinton has flaws, Donald Trump is temperamentally unfit to be President or to hold any public office in the United States.}}\\
 \bottomrule
 \end{tabular}
 \label{tbl:USrawdata}
\end{table}

\begin{table}[th!]
 \begin{tabular}{p{5em} p{3em} p{2em} p{38em}}
\toprule
Label & typical & count & responses \\
 \midrule
Neither (group 3) &Yes &3 & \textit{Donald Trump is a sham. He is only interested in his own popularity and power, and is a sexist, racist liar.}\\
 &Yes &3 & \textit{I am not a HillaryClinton fan, but I hope she's okay.} \footnotemark[9]\\
 &Yes &3 & \textit{ImWithHer because she cared enough to prepare. She's smart. Ready. And passionate about details because they matter.} \footnotemark[7]\\
 &Yes &2 & \textit{I am against Trump because he doesn't seem to have the  emotional stability required from someone that can launch nuclear weapons.}\\
 &Yes &2 & \textit{Trump repeatedly claimed he has``a foolproof way of winning the war with ISIS." How many will have to die before he shares it?} \footnotemark[10]\\
 &Yes &1 & \textit{America can't be compromised by leaders with conflicting loyalties.} \footnotemark[11]\\
 &Yes &1 & \textit{At this moment I think both of them have many problem with conducting the country.}\\
 &Yes &1 & \textit{Both candidates are not great, I don't vote but if I were to I would choose her as the lesser of two evils.}\\
 &Yes &1 & \textit{Never Trump Or Hillary! Neither are trustworthy or capable of transparent presidency promoting peace prosperity \& freedom. Im W/ Gary Johnson.}\\
 &Yes &1 & \textit{NeverTrump bc public persona of cruelty and predatory business practices, and a campaign conjuring xenophobia to answer with empty slogans.}\\
 &Yes &1 & \textit{WithHerBecause I'm exfootnotemarkd to have them BOTH back in the White House!} \footnotemark[12]\\
 &\textcolor{gray}{No} &\textcolor{gray}{4} & \textcolor{gray}{\textit{I think he doesn't have the minimal level of knowledge required to be capable of running a country.}}\\
 &\textcolor{gray}{No} &\textcolor{gray}{3} & \textcolor{gray}{\textit{ImWithHer because she cared enough to prepare. She's smart. Ready. And passionate about details because they matter.}} \footnotemark[7]\\
 &\textcolor{gray}{No} &\textcolor{gray}{2} & \textcolor{gray}{\textit{Donald Trump is a sham. He is only interested in his own popularity and power, and is a sexist, racist liar.}}\\
  &\textcolor{gray}{No} &\textcolor{gray}{2} & \textcolor{gray}{\textit{I'm not with her. I feel the bern. But we are stronger together.}}\\
 &\textcolor{gray}{No} &\textcolor{gray}{1} & \textcolor{gray}{\textit{America can't be compromised by leaders with conflicting loyalties.}} \footnotemark[11]\\
 &\textcolor{gray}{No} &\textcolor{gray}{1} & \textcolor{gray}{\textit{Donald trump is the outsider that America needs. Nationalism, not globalism.}}\\
 &\textcolor{gray}{No} &\textcolor{gray}{1} & \textcolor{gray}{\textit{Hillary has such an impressive history of working for us all!}} \footnotemark[6]\\
 &\textcolor{gray}{No} &\textcolor{gray}{1} & \textcolor{gray}{\textit{Hillary knows how it works.}}\\
 &\textcolor{gray}{No} &\textcolor{gray}{1} & \textcolor{gray}{\textit{I am against Trump because he doesn't seem to have the  emotional stability required from someone that can launch nuclear weapons.}}\\
 &\textcolor{gray}{No} &\textcolor{gray}{1} & \textcolor{gray}{\textit{I am not a HillaryClinton fan, but I hope she's okay.}} \footnotemark[9]\\
 &\textcolor{gray}{No} &\textcolor{gray}{1} & \textcolor{gray}{\textit{I believe that both are incapable of running this country.}}\\
 &\textcolor{gray}{No} &\textcolor{gray}{1} & \textcolor{gray}{\textit{I believe we need to break out of a two-party election.}}\\
 &\textcolor{gray}{No} &\textcolor{gray}{1} & \textcolor{gray}{\textit{I don't want neither of them to be my president}}\\
 &\textcolor{gray}{No} &\textcolor{gray}{1} & \textcolor{gray}{\textit{I'm with her because she is the most pro-science and intellectual candidate.}}\\
 &\textcolor{gray}{No} &\textcolor{gray}{1} & \textcolor{gray}{\textit{She believes America is already great and will improve it even more!}} \footnotemark[8]\\
 &\textcolor{gray}{No} &\textcolor{gray}{1} & \textcolor{gray}{\textit{She is the lesser of 2 evils.}}\\
 &\textcolor{gray}{No} &\textcolor{gray}{1} & \textcolor{gray}{\textit{Trump repeatedly claimed he has "a foolproof way of winning the war with ISIS. " How many will have to die before he shares it?}} \footnotemark[10]\\
 \bottomrule
 \end{tabular}
 %\label{tbl:USrawdata}
\end{table}

\footnotetext[1]{https://twitter.com/bfraser747/status/769360098183426048}
\footnotetext[2]{https://www.jta.org/2016/08/30/election-2016/why-do-floridas-orthodox-jews-support-trump-because-they-fear-clinton}
\footnotetext[3]{https://twitter.com/searfoss70/status/758286802964840448}
\footnotetext[4]{https://twitter.com/Dennisoflore/status/775232114530803712}
\footnotetext[5]{https://twitter.com/sportspickle/status/753612941195227136}
\footnotetext[6]{https://twitter.com/Varidienne/status/761704767076732928}
\footnotetext[7]{https://twitter.com/HoneyDemForce/status/779249860981747712}
\footnotetext[8]{https://twitter.com/Pammie3333/status/760580224073347072}
\footnotetext[9]{https://twitter.com/simplechefsue/status/775112417994952704}
\footnotetext[10]{https://twitter.com/MikeBates/status/773741717795147776}
\footnotetext[11]{https://twitter.com/ResisterDot/status/760710084175486976}
\footnotetext[12]{https://twitter.com/Varidienne/status/761705639303340032}

\clearpage

\subsection{Survey targeting graduates of a Faculty of Education}
This survey was conducted at the beginning of 2017, targeting the graduates of the Faulty of Education at a particular national university in Japan between 2000 and 2016.
We announced the survey by postal mail, sending invitations to 3,256 people who belong to this group.
% and 447 of postal mails were returned due to unknown address. 
All responses were collected through the web application.
We received 273 responses to Q1, 260 responses to Q2, and 258 responses to Q3. 
The survey was conducted in Japanese, and the questions and responses that appear in this paper are translations of the originals. (The translation was performed by the authors.) 
Typical responses and the group labels are listed in Table~\ref{tbl:labeling}. 

\begin{figure}[t!]
 \begin{center}
   \includegraphics[width=0.8 \textwidth]{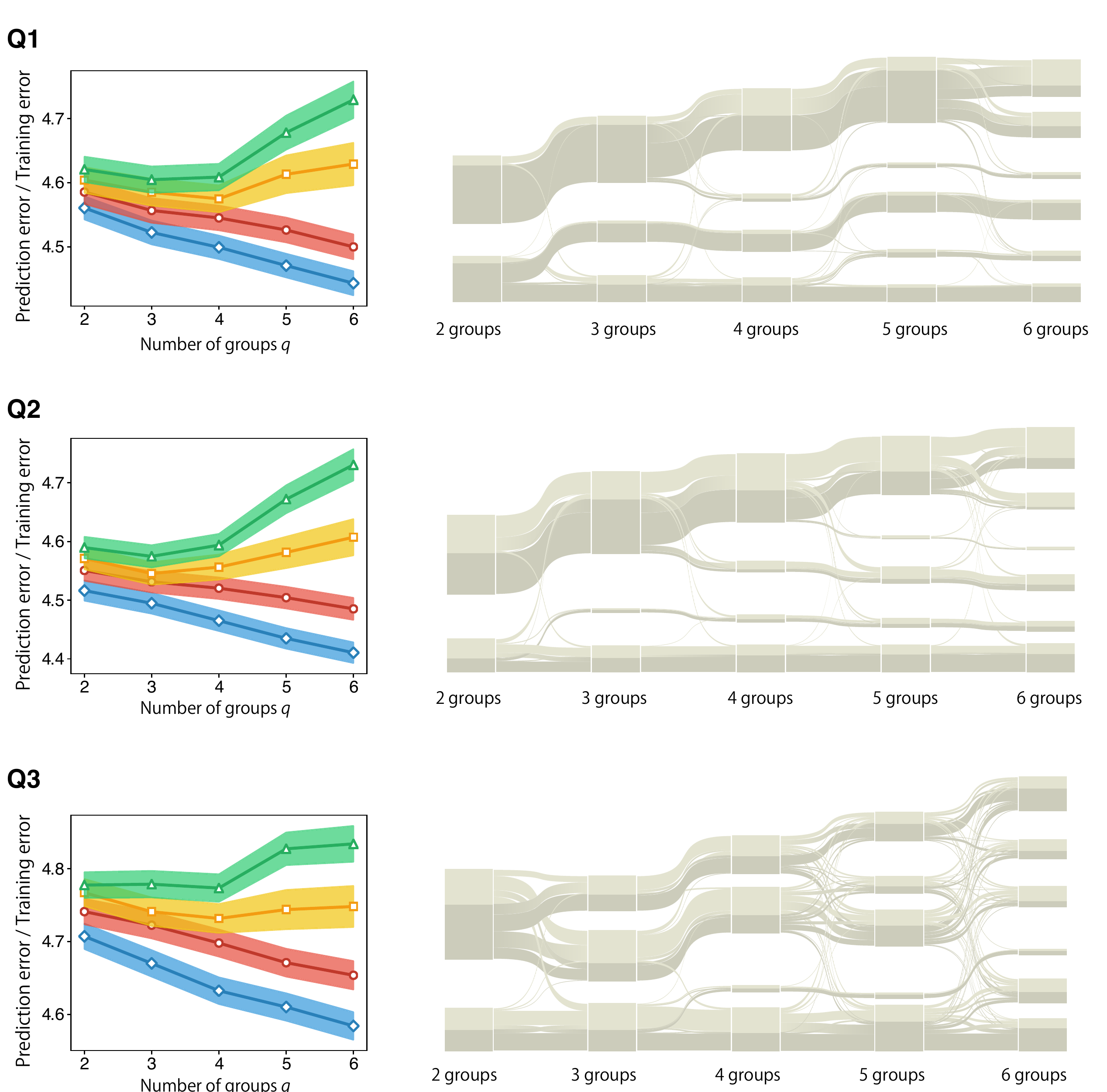}
 \end{center}
 \caption{
 Cross-validation errors (left) and alluvial diagrams (right) of the survey targeting graduates of a Faculty of Education. 
 The four cross-validation errors are plotted in the same panel using the same colour set as in Fig.~\ref{fig:LOOCV-US}. 
 In the alluvial diagrams, vertices with localised group assignments are depicted by dark colours.
}
\label{fig:alluvialKagawa}
\end{figure}

The cross-validation errors and the alluvial diagrams are displayed in Fig.~\ref{fig:alluvialKagawa}. 
For all questions, a unique minima can be observed for the Gibbs prediction errors (green) and its MAP estimates (yellow), while the Bayes prediction errors (red) seems to largely overfit.
Again, for all opinion graphs, we confirmed that the prediction errors for $q=1$ are significantly larger than those for $q \ge 2$.

For Q1, according to the cross-validation errors, the appropriate number of groups is three or four. 
The alluvial diagram exhibits a hierarchical structure only up to $q=4$ when we focus on the vertices with localised group assignments. 
In the main text, we focused on the partition into four groups.
The smallest group (group 4: `change of careers') appears to be robust because it exists under many different partitions. 
However, because this group does not exhibit an assortative structure, we consider the partition into three groups to also be possibly optimal.

\begin{figure}[t!]
 \begin{center}
   \includegraphics[width=\textwidth]{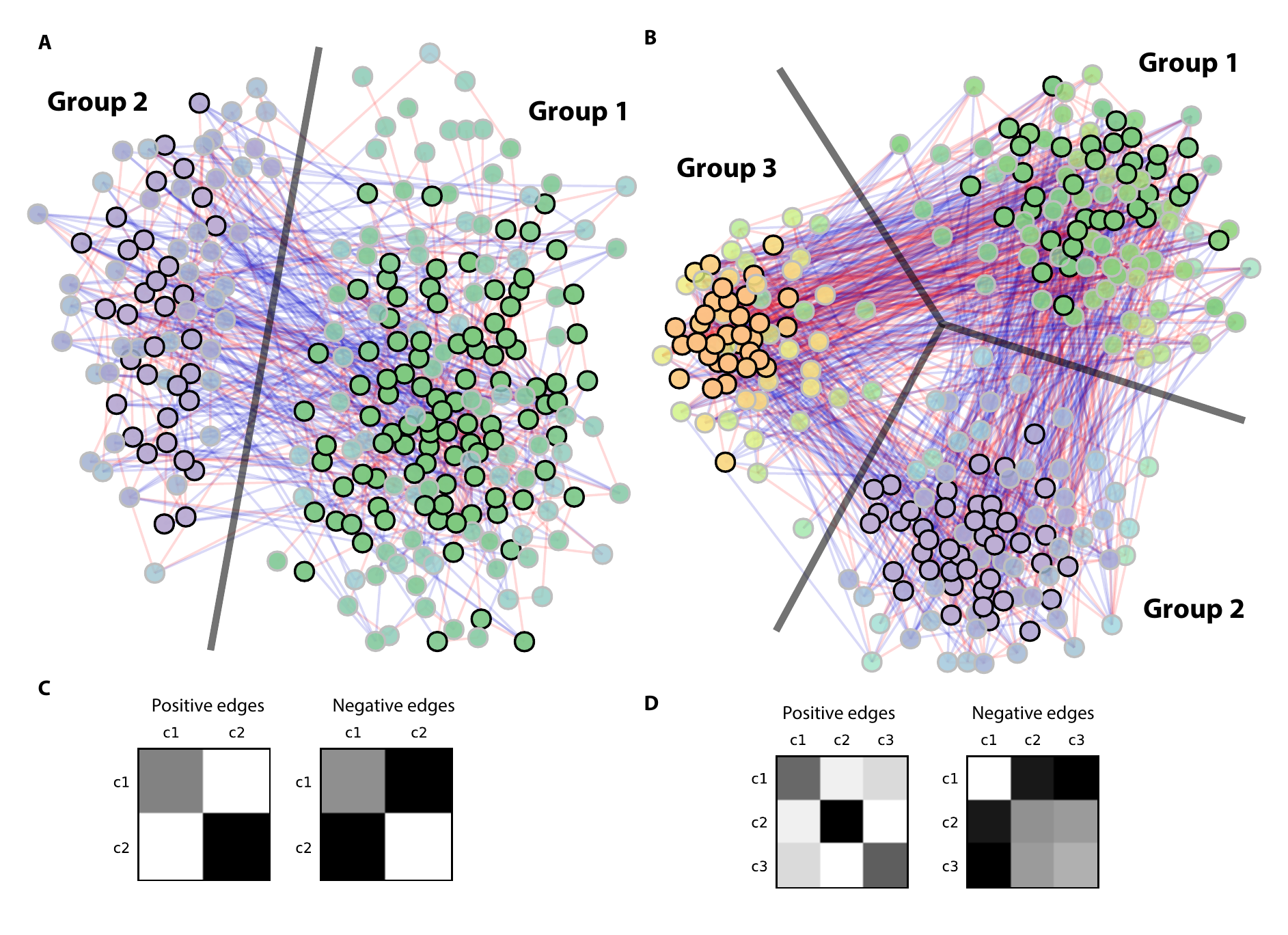}
 \end{center}
 \caption{
({\bf A}) Opinion graph for Q2. 
The positive and negative edges appear in red and blue, respectively.
The group-assignment probabilities for the two groups are indicated by  a colour gradient that varies between green (group 1) and purple (group 2).
The typical opinions that are selectively localised in a single group ($p_{\max} \ge 0.9$) have black borders, while others have grey borders.
({\bf B}) The opinion graph for Q3. 
This is plotted in the same manner as that for Q2, except that the group-assignment probabilities for the three groups are indicated by a colour gradient varying among green (group 1), purple (group 2), and yellow (group 3).
({\bf C}) Density matrices with respect to the positive and negative edges of Q2.
({\bf D}) Density matrices with respect to the positive and negative edges of Q3.
}
\label{fig:opinion_graphQ2Q3}
\end{figure}

For Q2, according to the cross-validation errors, the appropriate number of groups is two or three. 
The labels of the groups in the partition into two groups are listed in Table~\ref{tbl:labeling}. 
Group 2 (`working conditions') further splits into two groups in the case of a partition into three groups. 
It is indeed possible for a specific set of working conditions to be extracted as a group. 
However, after interpreting the partitions, we selected the partition into 2 groups as optimal for the present survey. 
The opinion graph for Q2 is displayed in Fig.~\ref{fig:opinion_graphQ2Q3} {\bf A}. 

For Q3, two, three, and four are possibly the most appropriate number of groups.
Because the hierarchical structure is lost in partitions into more than four groups, we conclude that the Bayes prediction error overfits. 
We selected the partition into three groups as the final choice because the distinction between the partitions into three and four groups is difficult to interpret.
The opinion graph for Q3 is presented in Fig.~\ref{fig:opinion_graphQ2Q3} {\bf B}. 

The density matrices of the positive and negative edges for Q2 and Q3 are depicted in Figs.~\ref{fig:opinion_graphQ2Q3} {\bf C} and {\bf D}, respectively. We have confirmed that assortative structures are learned for the positive edges, while disassortative structures are learned for the negative edges. 

\begin{table}
\caption{Labels on the opinion groups and the corresponding typical responses in the survey targeting graduates of a Faculty of Education.}
 \begin{tabular}{p{10em} @{\hspace{1em}} p{13em} @{\hspace{1em}} p{25em}}
% \begin{tabular}{p{10zw}@{\hspace{2zw}}l@{\hspace{1zw}}p{30zw}}
\toprule
Question  & Label  & Subset of typical responses \\
\midrule
Q1: `What is your career?'
 & 1. teachers & \textit{``Elementary school teacher'', ``Junior-high school teacher''}\\
 & 2. civil servants & \textit{``Working for a local government'', ``Police officer''}\\
 & 3. office workers & \textit{``Office worker in a local firm'', ``Banker'', ``Working in sales''}\\
 & 4. change of careers & \textit{``I used to work in a firm and left after having a baby. Now I'm working for a local government as having a temporary job at a kindergarten.''}\\
\midrule
Q2: `What is the reason you chose your current job?' 
 & 1. dream, rewarding, application of learned skills & 
 	\textit{``I've dreamed to be since childhood.'', 
	``It was my dream and is a rewarding job.'', 
	``I've wanted to be like the teacher I respected.'', 
 	``I can exercise what I learned in the faculty.''}\\
 & 2. working conditions & 
 	\textit{``I wanted to work for my home town. Also, local government employee is a stable job.'', 
	``I can work after getting married and having a baby.'', 
	``Good welfare and location (home town)''}\\
 \midrule
Q3: `What is the most valuable experience that you will take away with you from your time at the university?'
 & 1. academic knowledge and friendships established & 
	  \textit{``I could have advanced education.'', 
	  ``It's a treasure in my life to have good friends, and in-depth knowledge learned in university gave me a broader perspective.'',
	  ``I feel that I can see things from various angles since I met people with different opinions and learned several disciplines.'',
	  ``Liberal arts and academic knowledges gave me self-confidence.'',
	  ``All advices by professors and encounter with wonderful friends who were willing to be teacher like I did.''}\\
 & 2. certification and expertise & 
	  \textit{``Teaching certification.'',
	  ``Expertise, skills, and a way of thinking.'',
	  ``I'm not sure about the value what I learned, because I just wanted to receive the teaching certification.''
	  ``Education theory, because I didn't have enough time for reading books.''}\\
 & 3. self-discovery and social knowledge &
	  \textit{``I could have some time to think about myself and my future by meeting a lot of people.'',
	  ``I learned many things which couldn't learned in a classroom by conversations with a lot of people.''
	  ``Human relationships and the time of self-discovery'',
	  ``Communication skills and an inquiring mind'',
	  ``The skills of problem discovery and critical thinking and the importance of self-discovery''}\\
  \bottomrule
 \end{tabular}
\label{tbl:labeling}
\end{table}

% SECTION 4
%\section{Comparison between network-based and human-based classifications}
\section{Human-based classification}
% introduction [purpose of this section]
The most naive approach to handling raw data is to have annotators read all responses and classify them on the basis of their own subjective similarity criteria, i.e., human-based classification. 
This requires considerable human effort and each of the resulting classifications cannot be considered as a ``ground truth.'' 
In this section, we investigate the correlation between the opinion groups inferred from the opinion graph and those obtained from a human-based classification. 

%% detail of human-based classification
Human-based classification was performed by 5 annotators for Q1 and 8 annotators for Q2 and Q3. 
Each annotator first classified the responses into an arbitrary number of groups and then aggregated them into fewer than eight groups. 
Individuals possessing sufficient knowledge of the Faculty of Education are chosen as annotators to perform  meaningful classifications.

% definition of NMI rNMI ARI
We adopt two measures to compare two classifications: the normalized mutual information (NMI) \cite{NMI} and the adjusted rand index (ARI) \cite{ARI,Randindex}.
Let us write the classifications that we consider as $\ket{\sigma}_A$ and $\ket{\sigma}_B$. 
We use $a \in \{1,\dots,q_{A}\}$ and $b \in \{1,\dots,q_{B}\}$ to represent the group labels of the classifications $\ket{\sigma}_{A}$ and $\ket{\sigma}_{B}$, respectively.
The NMI is a measure commonly used to evaluate the similarity between classifications.
This measure is based on the mutual information $I(\ket{\sigma}_{A},\ket{\sigma}_{B})$ defined as follows;
\begin{align}
  I(\ket{\sigma}_{A},\ket{\sigma}_{B}) \equiv \sum_{a=1}^{q_{A}} \sum_{b=1}^{q_{B}} p(a,b) \log \frac{p(a,b)}{p(a) p(b)}.
\end{align}
Here, we have $p(a,b) \equiv N_{ab}/N$, $p(a) \equiv N_{a}/N$, and $p(b) \equiv N_{b}/N$, where $N$ is the total number of responses, $N_{ab}$ is the number of responses that belong to group $a$ in the classification $\ket{\sigma}_{A}$ and to group $b$ in the classification $\ket{\sigma}_{B}$, and $N_{a}$ and $N_{b}$ are the marginals of $N_{ab}$. 
The mutual information $I(\ket{\sigma}_A,\ket{\sigma}_B)$ measures the degree of agreement between two classifications. If two classifications are statistically independent, $I(\ket{\sigma}_A,\ket{\sigma}_B) = 0$, while having a positive value for similar classifications.
A problem is that mutual information tends to be large for fine-grained classification. 
To correct this tendency, the NMI is defined as  
\begin{align}
  \mathrm{NMI}(\ket{\sigma}_{A},\ket{\sigma}_{B}) \equiv \frac{2 I(\ket{\sigma}_{A},\ket{\sigma}_{B})}{H(\ket{\sigma}_{A}) + H(\ket{\sigma}_{B})},
\end{align}
where $H(\ket{\sigma}_{A})$ and $H(\ket{\sigma}_{B})$ are the entropy with respect to $\{ p(a) \}$ and $\{ p(b) \}$, respectively.

The ARI is the corrected-for-chance version of the Rand index, which also quantifies the agreement between two classifications.
A pair of responses are considered agreements if they belong to the same group in both classifications or if they belong to different groups in both classifications. 
Correcting the chance level of the Rand index, the ARI is defined as follows:
\begin{align}
  & \mathrm{ARI}(\ket{\sigma}_A,\ket{\sigma}_B)  = \notag \\
                                              &  \frac{\sum_{ab} \binom{N_{ab}}{2} - \left[  \sum_{a} \binom{N_{a}}{2}  \sum_{b} \binom{N_{b}}{2} \right]/ \binom{N}{2} }
{\frac{1}{2} \left[ \sum_{a} \binom{N_{a}}{2} + \sum_{b} \binom{N_{b}}{2}\right]  - \left[ \sum_{a} \binom{N_{a}}{2} \sum_{b} \binom{N_{b}}{2} \right] / \binom{N}{2}}.
\end{align}

% Three types of pair comparision
\begin{figure}[!htbp]
 \begin{center}
   \includegraphics[width=0.6 \textwidth]{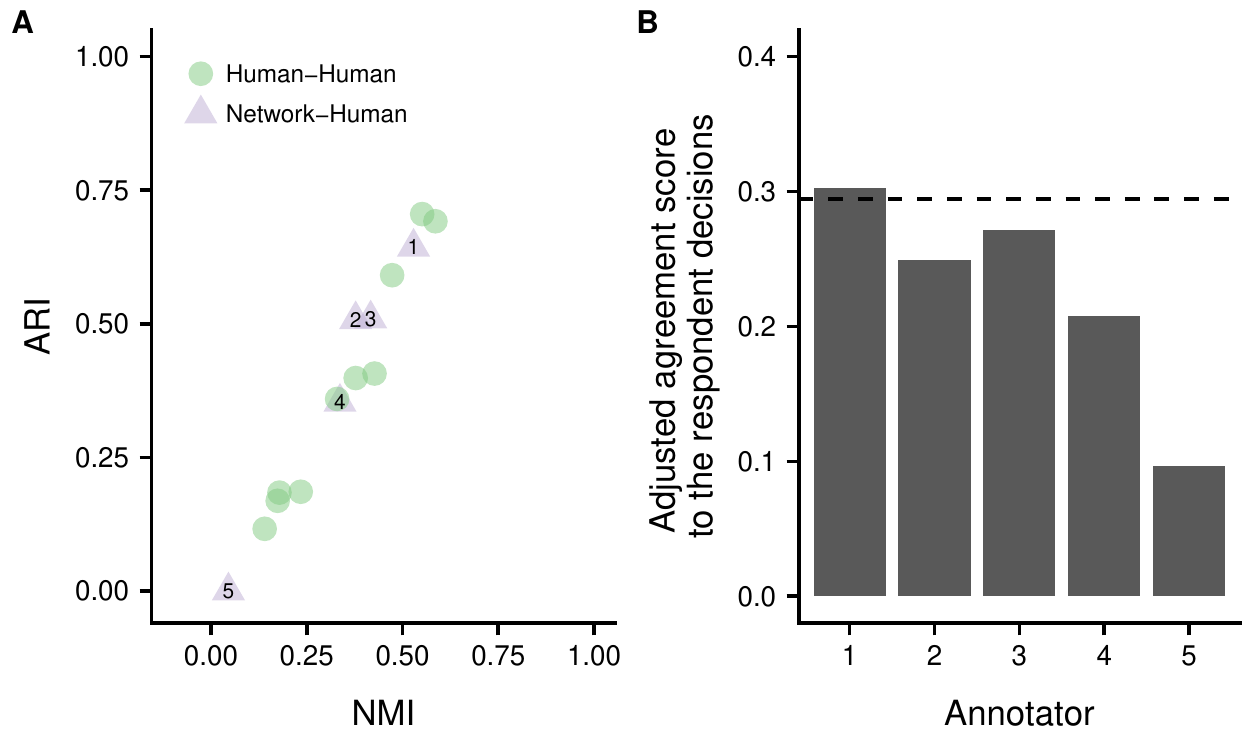}
 \end{center}
 \caption{
({\bf A}) Similarity measures, NMI and ARI, between different classifications of responses to Q1.
\textit{Network-Human} comparisons (circles) and \textit{Human-Human} comparisons (triangles) were performed by five annotators. The numbers identify the annotators.
({\bf B}) Adjusted agreement scores of the classifications performed by the annotators (bars) and the score of the network-based classification (horizontal dashed line). 
}
\label{fig:q1score}
\end{figure}

For the responses to Q1, we evaluated the NMI and ARI between the network-based and human-based classifications (Fig. \ref{fig:q1score}{\bf A}).
These data are labelled as \textit{Network-Human}.
In this comparison, we observe that four data points exhibit large values, while one data point is close to zero.
This indicates that the network-based classification is in good agreement with the classifications performed by the four annotators, even though the classification of one (annotator 5) is very different. 
In the same figure, we compare the classifications among the annotators (\textit{Human-Human}).

It is possible that naive human-based classifications would be inconsistent with the decisions made by the respondents, i.e., the edge labels obtained in our approach.
If this is indeed the case, dissimilarity will inevitably appear in the \textit{Network-Human}.
Such a dissimilarity would not be judged as indicating a flaw in our approach. 
To verify the consistency between the classifications performed by the annotators and the decisions made by the respondents, we measured the agreement score (AS) of the human-based classification $\ket{\sigma}_{A}$, defined as follows:
\begin{align}
 \text{agreement score (AS)}(\ket{\sigma}_{A}) = \frac{M_{+(=)} + M_{-(\neq)}}{M},
\end{align}
where $M_{+(=)}$ is the number of positive edges that connect the responses in the same group in $\ket{\sigma}_{A}$, $M_{-(\neq)}$ is the number of negative edges that connect the responses in different groups in $\ket{\sigma}_{A}$, and $M$ is the total number of both types of edges. 
Next, we introduce a corrected-for-chance version of the agreement score, which we refer to as the ``adjusted agreement score''. This is defined as follows:
\begin{align}
 \text{adjusted agreement score}(\ket{\sigma}_{A}) &= \text{AS}(\ket{\sigma}_{A}) - \bracket{ \text{AS}(\ket{\sigma}_{C}) },
\end{align}
where $\ket{\sigma}_{C}$ is a randomised classification of $\ket{\sigma}_{A}$, in which the group assignments are randomised while maintaining the group size distribution, and $\bracket{\cdots}$ is the ensemble average over the realisations of $\ket{\sigma}_{C}$.
Figure \ref{fig:q1score}{\bf B} presents the adjusted agreement scores for classifications performed by the annotators for Q1. 
We observe that the classification of annotator 5, which exhibits significant disagreement with the network-based classification, also exhibits a low adjusted agreement score. 
Therefore, this classification must be treated as an outlier.

\begin{figure}[!htbp]
 \begin{center}
   \includegraphics[width=0.6 \textwidth]{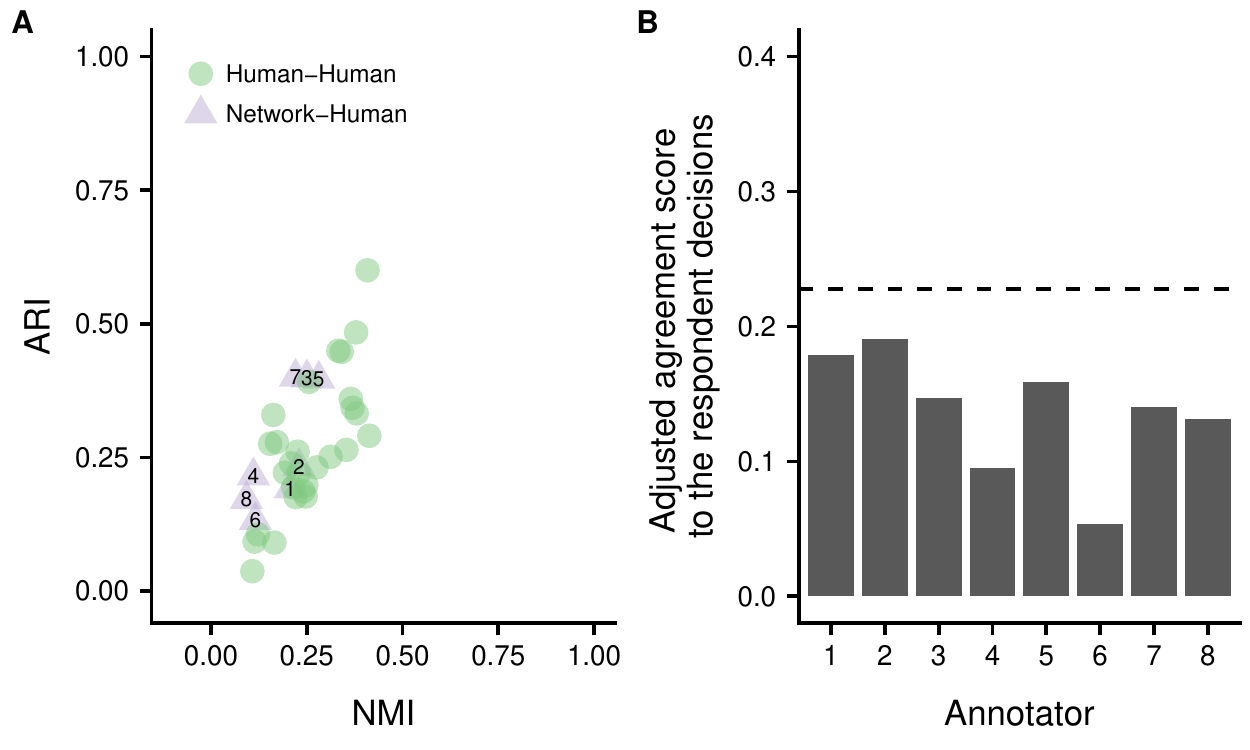}
 \end{center}
 \caption{Comparisons analogous to those in Fig.~\ref{fig:q1score} for Q2.
The human-based classifications were performed by eight annotators.}
\label{fig:q2score}
\end{figure}

\begin{figure}[!htbp]
 \begin{center}
   \includegraphics[width=0.6 \textwidth]{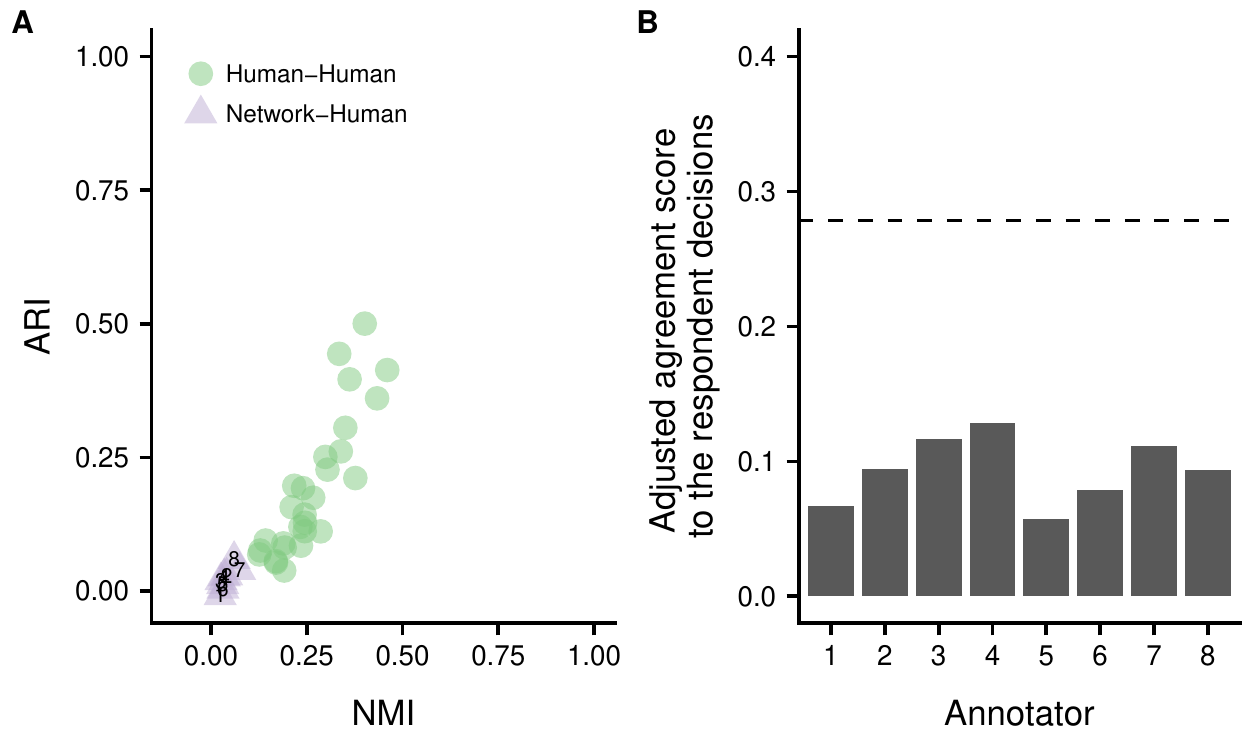}
 \end{center}
 \caption{Comparisons analogous to those in Fig.~\ref{fig:q1score} for Q3.
The human-based classifications were performed by eight annotators.
}

\label{fig:q3score}
\end{figure}

Analogous comparisons for Q2 and Q3 are illustrated in by Figs.~\ref{fig:q2score} and \ref{fig:q3score}. 
For Q2, the situation is very similar to that of Q1, except that the values of the NMI, ARI, and adjusted agreement scores are smaller than those of Q1.
For Q3, we can observe from Fig.~\ref{fig:q3score}{\bf A} that the NMIs and ARIs for the \textit{Network-Human} comparisons are systematically lower than those for the \textit{Human-Human} comparisons. 
Moreover, Fig.~\ref{fig:q3score}{\bf B} indicates that for every annotator, agreement of the human-based classification with the decisions made by the respondents is significantly poorer than the network-based classification (horizontal line). 
We conclude that the classifications determined by the annotators are simply inconsistent with the data on the network.

%%%
%%% Text Analysis
%%%
%\section{Comparison between network-based and NLP-based classifications}
\section{Natural language processing-based classifications}
Feature extraction from text data is a technique that is widely applicable whenever datasets are written in natural languages. 
Such a field is sometimes referred to as the natural language processing (NLP). 
Classification based on the text analysis is in fact an alternative approach to classify the collected opinion texts. 
In this section, we compare two commonly used NLP techniques --- the latent Dirichlet allocation (LDA) \cite{LDA_Blei2003} and Word2Vec \cite{word2vec_NIPS2013} --- with the network-based and human-based classifications.

Before we move on to the detailed analysis, it is important to note that these NLP-based classifications identify opinion groups on the basis of word similarity in the opinion texts. 
This is distinct from the network-based classification in which opinion groups are determined on the basis of the semantic similarity. 
Therefore, although the aims of these approaches are analogous to each other, they serve different objectives and their results can be inconsistent. 
In particular, these differences are critical for the survey analysis, in which semantic relations are really significant. 
Whereas the NLP methods can be successful in identifying semantic relationships in some cases \cite{word2vec_NIPS2013}, our approach works by construction. 

\subsection{Methods of NLP-based classification}
The preprocessing has to be conducted carefully in the NLP, as it may critically affect the overall performance. 
For a given set of the documents, i.e., the set of opinion texts in the present case, we first break each document into a set of words (or morphemes). 
This preprocess is known as the tokenization. 
To this end, we need to identify the language(s) and lexical categories. 
We also need to define the stop words and lower and upper cutoff with respect to the co-occurrence frequencies and lengths of the words. 
Moreover, we need a dictionary to lemmatize each word. 
Hereafter, we assume that the dataset is tokenized.

\subsubsection{Topic modelling based on the LDA}
A classical approach to the text analysis is the LDA. 
In the LDA, we consider a bipartite graph that has word vertices and document vertices. 
Again, the document vertices are the opinion vertices in the present case and the word vertices are the whole set of tokenized word elements. 
A pair of a document vertex and a word vertex is connected by an edge when the word is included in the document. 
The task of the LDA is to identify the groups of documents as well as the groups of words. 
For the implementation of the LDA, we used the code distributed at \cite{gensim_rehurek_lrec}. 
Hereafter, we show the results of a specific number of groups among the ones we have tried.

\subsubsection{Classification based on the word embedding}
Another NLP-based approach is the word embedding. 
A number of word embedding methods have been proposed during the several years, e.g., Word2Vec, Glove \cite{pennington2014glove}, and fastText \cite{fastText_joulin2017bag}, to name a few. 
Here, we consider a document classification using the Word2Vec. 

We again use the tokenized words for each document. 
The Word2Vec maps each word to a high-dimensional vector such that the distance between a pair of vectors represents the similarity between these words. 
These vectors are determined by training a corpus so that the vector similarity correctly represents the frequency of word co-occurrence in the corpus. 
To obtain a vector corresponding to a document, we take the average over the word vectors within the document. 
For the set of document vectors, we determine the classification by the k-means algorithm \cite{KmeansPlusPlus}. 

To perform the mapping, we adopted pretrained word vectors of a large corpus for each language: \cite{GoogleCorpusURL} for English and \cite{JapaneseCorpusURL} for Japanese. 
Therefore, the similarity among the words is a similarity in a generic sense and does not reflect the background of each survey. 
Moreover, the corpus must contain the words in the dataset. 
In this sense, when we expect the word embedding to perform well, we need to correctly impose the domain knowledge in the analysis. 

We also performed the same analysis by using the dataset itself as a corpus to be trained. 
We do not need to impose a domain knowledge in this case. 
A problem of this approach is that the word similarity may not be correctly captured when the dataset is small. 
Note that the fact that the dataset of opinion texts is small is not equivalent to the fact that the opinion graph is small; it may happen that unique opinions are few even if a large number of respondents participates in the survey. 
We do not show the results of this approach because we could not confirm a dramatic difference from the above results. 

It is also possible to fine-tune the word embedding by further training the dataset on the basis of the pretrained word vectors. 
However, none of the datasets considered in this paper appear to be large enough so that a fine-tuning would cause a significant improvement. 
%However, none of the datasets considered in this paper does not appear to be large enough so that a fine tuning causes a significant improvement. 

\subsection{Application to the datasets}
\subsubsection{Poll on the 2016 US presidential election}
\begin{figure}[!htbp]
 \begin{center}
   \includegraphics[width=0.9 \textwidth]{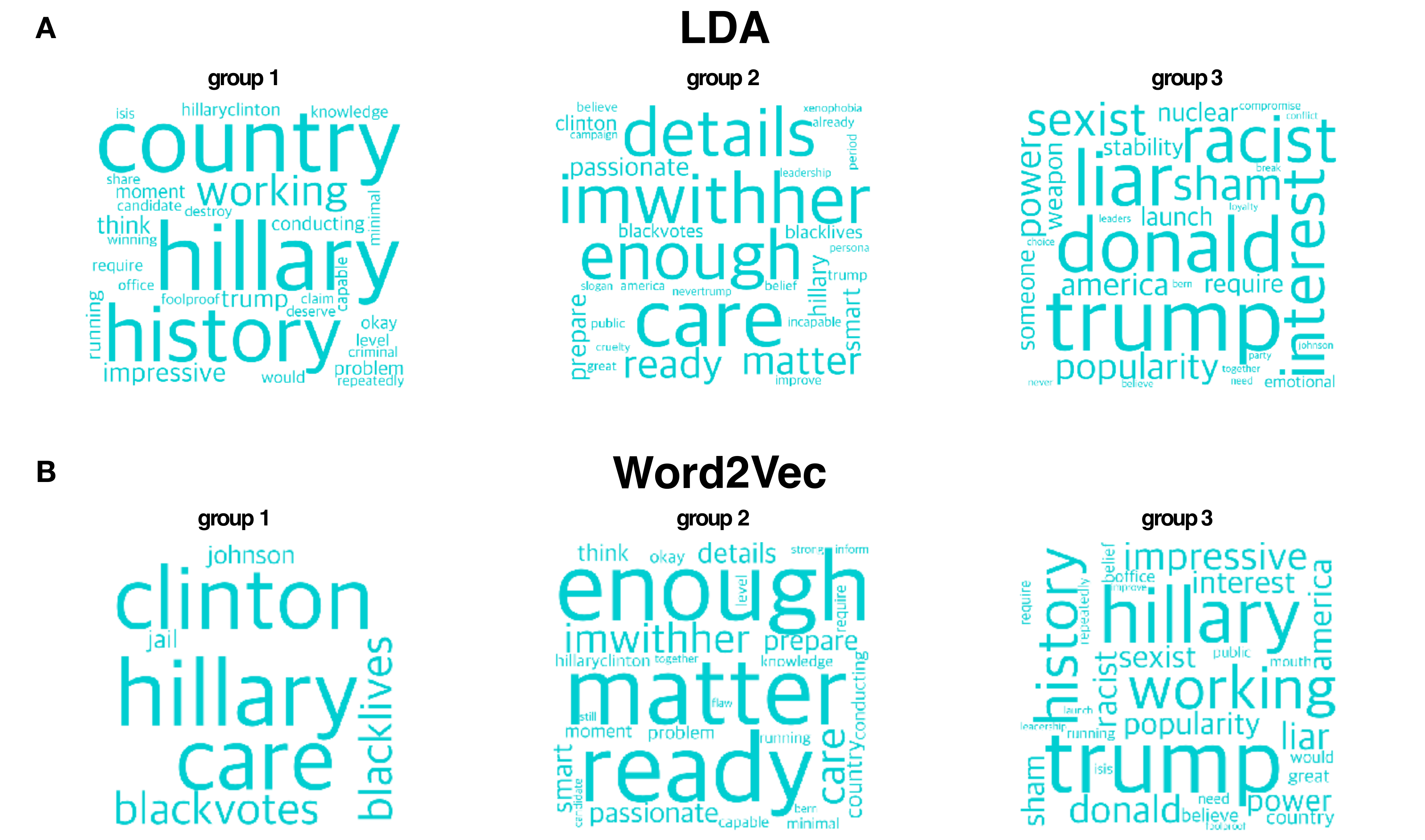}
 \end{center}
 \caption{WordClouds of the groups identified by the LDA (A) and Word2Vec (B). }
\label{fig:wordcloud-US}
\end{figure}

We first show the results for the US presidential election dataset. 
Figure \ref{fig:wordcloud-US} shows the WordCloud corresponding to each group. 
The LDA directly extracts a set of words as a group and each word has its own weight. 
For the Word2vec, based on the result of document classification, we counted the word frequency within a group to generate the WordCloud. 
In both cases, we can see that the opinion texts are classified into two groups (topics 1 and 2) related to Hillary Clinton and one group (topic 3) related to Donald Trump. 
However, we can hardly capture the semantics from these WordClouds. 
Note that, for the LDA, group 3 is misleading because it contains words that are used to describe Hillary Clinton. 
Such an inconvenience can easily occur because the names of both candidates are frequently mentioned by both types of supporters.

\begin{figure}[!htbp]
 \begin{center}
   \includegraphics[width=0.4 \textwidth]{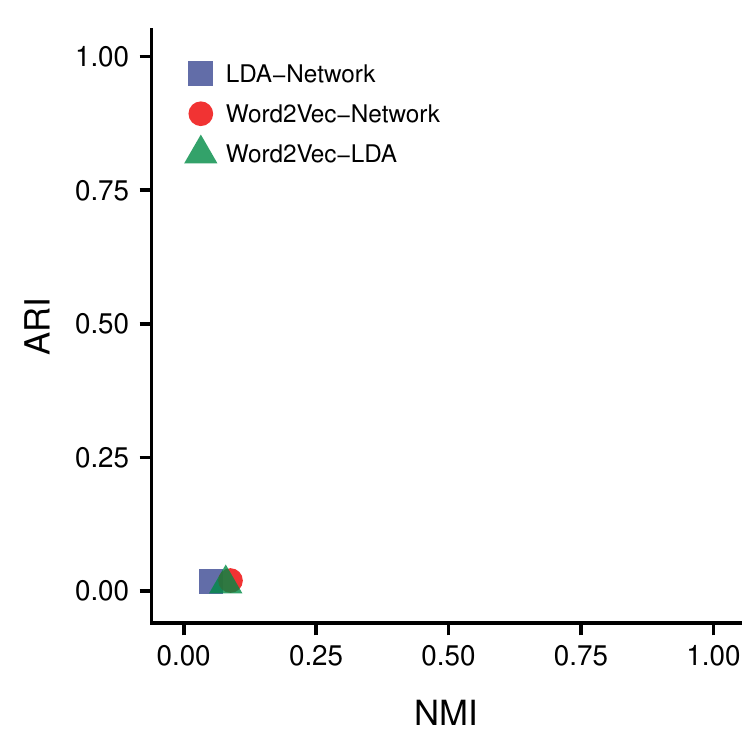}
 \end{center}
 \caption{Similarity measures, NMI and ARI, between NLP-based and network-based classifications of responses to Poll on the 2016 Us presidential election.  }
\label{fig:score_nlp_us}
\end{figure}

Let us then show the comparison of the document classifications. 
Analogously to Fig.~\ref{fig:q1score}, the results of LDA and Word2Vec with the network-based classification are compared in Fig.~\ref{fig:score_nlp_us}. 
Although the similarities among these methods are nonzero, they are considerably low. 
Even the similarity between the LDA and Word2Vec are close to zero.

\subsubsection{Survey targeting graduates of a Faculty of Education}
We next show the result for the faculty survey datasets. 
%We classify each dataset into the same number of groups as we did for the network-based classification. 
The representative words for each group are apparently similar regardless of the LDA and Word2Vec for Q1 and Q2. 
For Q1, one group contains the words ``teacher'' and ``primary school.'' 
The other groups contain ``graduation,'' ``company,'' and ``employed'' as representative words. 
For Q2, we obtained a group with ``work'' and ``interest,'' and a group with ``dream.'' 
Finally, for Q3 with LDA, we obtained a group with ``perspective'' and ``people,'' a group with ``knowledge'' and ``profession,'' and a group with no particular representative words. 
In contrast, we obtained no representative word for all groups with Word2Vec. 
Overall, when we focus on the keywords, we can see some correspondence with the results of network-based classification. 

\begin{figure}[!htbp]
 \begin{center}
   \includegraphics[width=0.7 \textwidth]{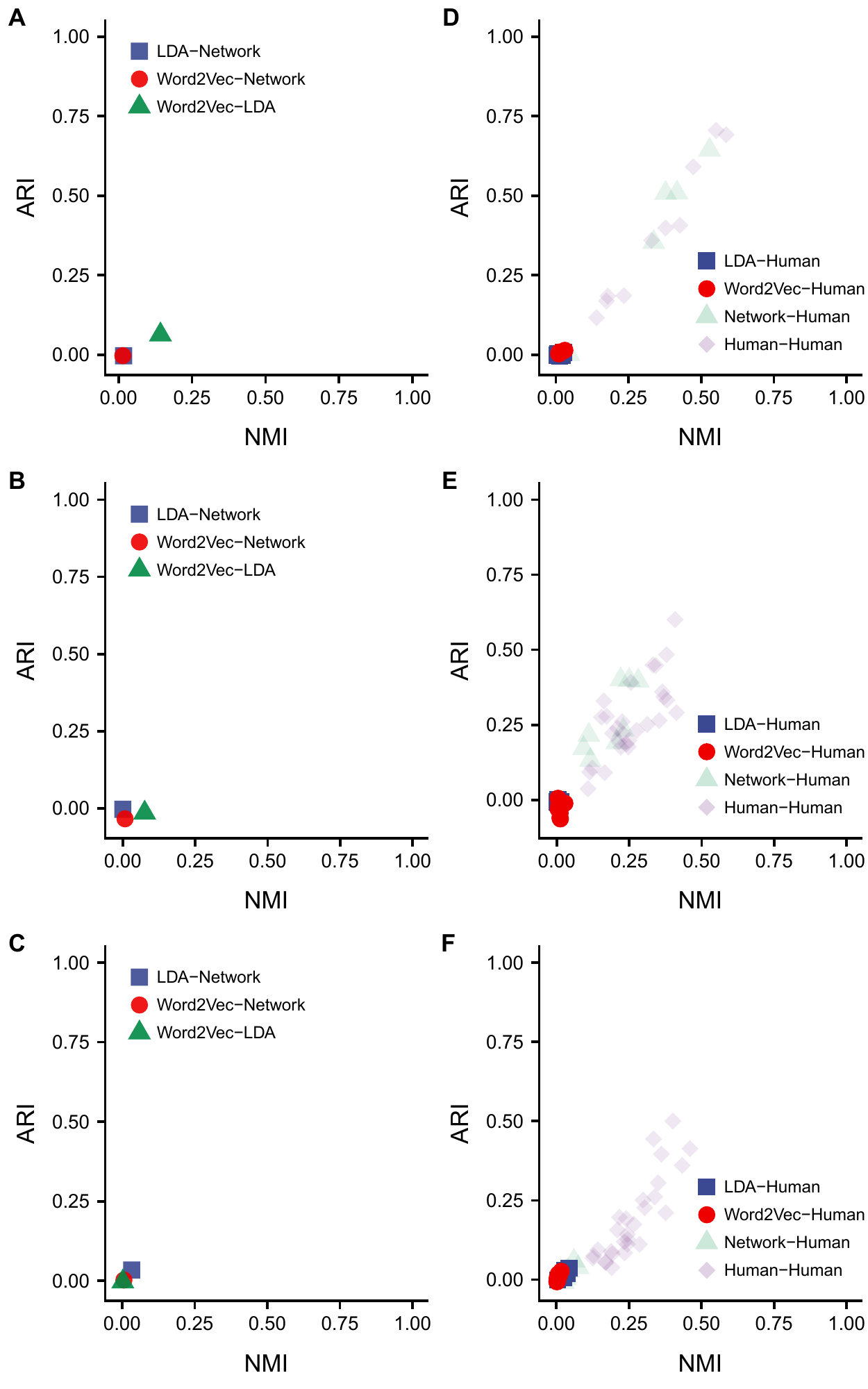}
 \end{center}
 \caption{
Comparison of different classifications for the faculty survey dataset. 
(Left column) Similarity measures, NMI and ARI, between NLP-based and network-based classifications of responses to Q1({\bf A}), Q2 ({\bf B}), and Q3 ({\bf C}), respectively. 
(Right column) Similarity measures, NMI and ARI, between NLP-based and human-based classifications of responses to Q1({\bf D}), Q2 ({\bf E}), and Q3 ({\bf F}), respectively. 
For the reference purpose, the results in Figs.~\ref{fig:q1score}, \ref{fig:q2score}, and \ref{fig:q3score} are included again. 
Each of the human-based classifications is compared with the NLP-based classifications (the LDA and Word2Vec). 
As we mentioned, the human-based classifications were performed by five annotators for Q1, while they were done by eight annotators for Q2 and Q3. 
 }
\label{fig:score_nlp_kagawa}
\end{figure}

The comparisons among the results of document classification is again conducted. 
This time, in addition to the comparison with the network-based classification (Figs.~\ref{fig:score_nlp_kagawa}A, B, and C), we also compare it with the human-based classification (Figs.~\ref{fig:score_nlp_kagawa}D, E, and F). 
The similarities of the NLP-based classifications to the network-based and human-based classifications are considerably lower than the similarity between the network-based and human-based classifications. 
This implies that, although a few representative opinions are consistently classified, there exist many opinions that are classified inconsistently. 

\subsection{Combination of the NLP-based and network-based classifications}
It is possible to combine the result of the NLP-based and network-based classifications. 
In fact, there is an approach \cite{Hu2014ITM} in which human annotations can be used as the metadata of the NLP. 
Similarly, for example, we can treat the results of NLPs as the metadata of the opinion vertices. 
However, such postprocessing should be applied carefully for the present purpose because, as we mentioned at the beginning of this section, the NLP-based and network-based classifications obey different principles. Therefore, the similarities and differences imposed by the texts may erroneously bias the semantic similarity between opinions.

%However, such postprocessing can be disastrous. 
%Again, as we mentioned at the beginning of this section, the NLP-based and network-based classifications obey different principles. 
%Therefore, the similarities and differences imposed by the texts may erroneously bias the semantic similarity between opinions.
%As we have observed in the third group in Fig.~\ref{fig:LDA-US}, a critical semantic difference that could be resolved in the original opinion graph may not be resolved after a NLP-based postprocessing.

A much better way to combine the NLP-based and network-based classifications is to include the NLP-based classification in our proposed survey process. 
For example, instead of the uniform sampling of opinions, we can perform biased sampling on the basis of the NLP-based classification. 
The assessment on biased sampling is left as future work.

%%
%% Regression
%%
\section{Logistic regression analysis on the inferred groups in the opinion graphs}

\begin{table}[!htbp] \centering 
  \caption{Logistic regression analysis is used to predict whether a respondent is a school teacher on the basis of his/her responses to the career motivation (Q2) and value of education (Q3) questions.} 
  \label{tbl:regression} 
\begin{tabular}{@{\extracolsep{5pt}}lD{.}{.}{-2} D{.}{.}{-2} D{.}{.}{-2} } 
\\[-1.8ex]\hline 
\hline \\[-1.8ex] 
 & \multicolumn{3}{c}{\textit{Dependent variable:}} \\ 
\cline{2-4} 
\\[-1.8ex] & \multicolumn{3}{c}{Dummy for the `teachers' group in Q1} \\ 
\\[-1.8ex] & \multicolumn{1}{c}{(1)} & \multicolumn{1}{c}{(2)} & \multicolumn{1}{c}{(3)}\\ 
\hline \\[-1.8ex] 
 Dummy for the `dream, rewarding, application of learned skills'\\ group in Q2 & 1.47^{***} & 1.56^{***} &  \\ 
  & \multicolumn{1}{c}{(0.83$, $2.10)} & \multicolumn{1}{c}{(0.94$, $2.18)} &  \\ 
  & & & \\ 
 %\parbox{27em}{Dummy for the `academic knowledge and friendships established' group in Q3} & 1.29^{***} &  & 1.43^{***} \\ 
 Dummy for the `academic knowledge and friendships established' group in Q3 & 1.29^{***} &  & 1.43^{***} \\ 
  & \multicolumn{1}{c}{(0.54$, $2.05)} &  & \multicolumn{1}{c}{(0.71$, $2.15)} \\ 
  & & & \\ 
 Dummy for the `certification and expertise' group in Q3 & 0.69^{*} &  & 0.68^{*} \\ 
  & \multicolumn{1}{c}{(-0.06$, $1.44)} &  & \multicolumn{1}{c}{(-0.04$, $1.39)} \\ 
  & & & \\ 
 Constant & -1.61^{***} & -0.91^{***} & -0.61^{**} \\ 
  & \multicolumn{1}{c}{(-2.38$, $-0.84)} & \multicolumn{1}{c}{(-1.44$, $-0.37)} & \multicolumn{1}{c}{(-1.18$, $-0.03)} \\ 
  & & & \\ 
\hline \\[-1.8ex] 
Observations & \multicolumn{1}{c}{233} & \multicolumn{1}{c}{233} & \multicolumn{1}{c}{233} \\ 
Log Likelihood & \multicolumn{1}{c}{-140.88} & \multicolumn{1}{c}{-146.81} & \multicolumn{1}{c}{-151.81} \\ 
Akaike Inf. Crit. & \multicolumn{1}{c}{289.77} & \multicolumn{1}{c}{297.63} & \multicolumn{1}{c}{309.63} \\ 
\hline 
\hline \\[-1.8ex] 
  \multicolumn{4}{r}{\textit{Note:}   95 percent confidence interval in parentheses. $^{*}$p$<$0.1; $^{**}$p$<$0.05; $^{***}$p$<$0.01} \\ 
\end{tabular} 
\end{table}

After the opinion groups have been identified, we perform categorical data analysis analogously in the case of multiple-choice surveys. 
In the main text, we presented bar charts and a Sankey diagram in Fig.~3, which presents the results. These results suggest correlations among the factors queried on the careers  of respondents (Q1), their career motivation (Q2), and the value of their education (Q3).

To investigate these correlations more accurately, we performed a logistic regression analysis and summarized the results in Table \ref{tbl:regression}. 
In this analysis, a dependent variable is whether a graduate is a school teacher or not, and independent variables are the dummy variables with respect to the opinion groups in Q2 and Q3.
In the regression analysis, we used the following three models:
\begin{align}
  \mathrm{logit}(Y_i)    
                      &= \beta_0 + \beta_{Q2} X_{Q2,i} + \beta_{Q3-1} X_{Q3-1,i} + \beta_{Q3-2} X_{Q3-2,i} \tag{model 1}\\
                      &= \beta_0 + \beta_{Q2} X_{Q2,i} \tag{model 2} \\
                      &= \beta_0 + \beta_{Q3-1} X_{Q3-1,i} + \beta_{Q3-2} X_{Q3-2,i} \tag{model 3},
\end{align}
where $Y_i$, $X_{Q2,i}$, $X_{Q3-1,i}$, $X_{Q3-2,i}$ are dummy variables that represent the `teachers' group for Q1, the `dream, rewarding, application of learned skills' group for Q2, and the `academic knowledge and friendships established' group and the `certification and expertise' group for Q3, respectively.
The results demonstrate that the respondents in the `teachers' group for Q1 are likely to be in the `dream, rewarding, application of learned skills' group for Q2 and in the `academic knowledge and friendships established' group for Q3.
The factor `certification and expertise' for the value of education queried in Q3 is positive, but not as significant as the factor `academic knowledge and friendships established' in this survey.
The factors `working conditions' for career motivation (Q2) and `self-discovery, social knowledge' for the value of education (Q3) have negative impacts on the `teachers' group, which is reflected in the value of the constant terms in models 2 and model 3.

\subsection{Utility of raw texts}
We can obtain even richer information from the raw texts than would be obtained through an ordinary categorical data analysis under multiple-choice questions. 
For example, while regression analysis reveals that the graduate who belongs to `dream, rewarding, application of learned skills' tends to belong to `teachers', we can further observe that there are some differences in the expressions between teachers and non-teachers.
For teachers, their dreams and desires were formed since childhood and often motivated by school teachers as role models.
On the other hand, non-teachers were observed to be typically motivated during high-school and undergraduate study and some of them were motivated by their parents.
Moreover, popular words ``rewarding'' and ``yearning to be'' in the teacher opinions are rarely observed in the non-teacher opinions. 
Such data can help us reveal in detail the reasons individuals choose to be teachers based on the result of the regression analysis.

%\info{Todo : check the citation ctyle}
%\bibliographystyle{apsrev}
\onecolumngrid
\bibliographystyle{naturemag}
%\bibliography{ref}

\end{document}